\documentclass[twocolumn,showpacs,prb]{revtex4}

\usepackage{dcolumn}
\usepackage{bm}
\usepackage{amsmath,amsthm}
\usepackage{amssymb}
\usepackage{graphicx}
\usepackage{color}
\usepackage{epic}
\usepackage{enumerate}

\newcommand{\dps}{\displaystyle }
\newcommand{\ri}{\text{i}}

\newcommand{\cT}{{\cal T}}

\newcommand{\tr}{\textrm{Tr}}
\newcommand{\OO}{\textrm{O}}

\newlength{\minipagewidth}
\newtheorem{algo}{Algorithm}
\newcommand{\bookbox}[1]{
\par\bigskip\noindent
\framebox[0.48\textwidth]{
\begin{minipage}{\minipagewidth}
{#1}
\end{minipage} } \par\medskip }

\begin{document}

\preprint{APS/}

%
%
  
\title{Thermal transport in isotopically disordered carbon nanotubes}

\author{Gabriel Stoltz}
\altaffiliation[Also at ]{
  IMPMC, Universit\'es Paris 6 et 7, CNRS, IGPG, 140 rue de Lourmel,
  75015 Paris, France
}
\affiliation{
  Universit\'e Paris Est, CERMICS,
  Project-team Micmac, INRIA-Ecole des Ponts, 
  6 \& 8 Av. Pascal, 77455 Marne-la-Vall\'ee Cedex 2, France
}
\author{Michele Lazzeri}
\author{Francesco Mauri}
\affiliation{
  IMPMC, Universit\'es Paris 6 et 7, CNRS, IGPG, 140 rue de Lourmel,
  75015 Paris, France
} 

\date{\today}

\begin{abstract}
We present a study of the phononic thermal conductivity of
isotopically disordered carbon nanotubes. 
In particular, the behavior of the thermal conductivity as a
function of the system length is investigated, using Green's function techniques
to compute the transmission across the system.
The method is implemented using linear scaling algorithms, which
allows us to reach systems of lengths up to $L=2.5~\mu$m
(with up to 400,000 atoms). 
As for 1D systems, it is observed that the conductivity diverges with the
system size $L$.  
We also observe a dramatic decrease of the thermal
conductance for systems of experimental sizes (roughly 80\% at room
temperature for $L = 2.5$~$\mu$m), when a large fraction of isotopic disorder is introduced.
The results obtained with Green's function techniques are compared to
results obtained with a Boltzmann description of
thermal transport. There is a good agreement between both approaches for systems of
experimental sizes, even in presence of Anderson localization. 
This is particularly interesting since the computation of the
transmission using Boltzmann's equation is much less computationally
expensive, so that larger systems may be studied with this method.
\end{abstract}

\pacs{
65.80.+n, 
61.46.Fg, 
63.22.-m, 
63.20.kp 
}

\maketitle

%
%

\section{Introduction}

Carbon nanotubes (CNTs) are very interesting materials for nanoscale
electronic devices due to their outstanding 
mechanical and electrical 
(depending on their chirality, CNTs can be semiconducting
or metallic) properties. Recently, it was also
discovered that CNTs have very good thermal properties, as 
measured experimentally for individual single walled
carbon nanotubes\cite{YSYLM05} or as estimated by computer
simulations (see the references in Section~\ref{sec:heat_CNT}). 
At room temperature, the thermal conductance of 
carbon nanotubes seems to be dominated by the phononic contribution, 
even for metallic carbon nanotubes.\cite{HWPZ99,YNW07}

Thermal properties are usually investigated using Fourier's law.
For nonequilibrium steady
states where the system is put in contact with two reservoirs at
different temperatures, there is a net energy flow from the hotter to
the colder reservoir.
The heat current density $J$ is proportional to the temperature gradient
\begin{equation}
  \label{eq:Fourier}
  {\bf J} = \kappa \nabla T,
\end{equation}
$\kappa$ being the thermal conductivity (a tensor, in general).
Denoting by $\Delta T$ the temperature
difference between the reservoirs, and by $L$ the system size in the
direction of the temperature gradient,
\[
\kappa = \frac{|{\bf J}| L}{\Delta T},
\]
provided the temperature profile is linear.
For usual three dimensional materials, the thermal conductivity does
not depend on the system size, and so, it is a well-defined
thermodynamical quantity.
The situation is different for one-dimensional (1D) 
systems, for which the thermal flux is
related to the transmission function of phonons from one
reservoir to the other. For defect free periodic 
one-dimensional harmonic systems,
there is no phonon scattering mechanism. Those
systems can sustain a current which does not
depend on the system's length,\cite{RLL67} 
so that $|{\bf J}|/\Delta T$ is constant. The thermal conductivity 
therefore diverges as $L$, and is not well-defined.
In general, one dimensional (1D) systems
in which scattering processes can take place 
should exhibit an intermediate scaling 
$|{\bf J}|/|\nabla T|\sim L^\alpha$ with 
$0 < \alpha < 1$, in which case the thermal conductivity 
is again not well-defined.

It is believed that CNTs 
should have a behavior reminiscent of 1D systems, although such
claims should be backed up by more systematic studies.
The dependence of the CNT conductivities on the system length should
therefore be a major and primary concern in any study of the thermal conductance.
Only very few experimental studies on the length dependence of the
thermal conductance of carbon nanotubes are available to our knowledge
(see for instance Ref.~\onlinecite{WTZZZ07} and
Ref.~\onlinecite{COGMZ08}), and numerical results are still
rare (see the references in Section~\ref{sec:heat_CNT}).
More importantly, Fourier's law, which is not valid in general for 1D
systems, is sometimes assumed to hold to
interpret experimental measurements in order to extract 
conductivities.\cite{PMWGD06} 

In order to have a well-defined conductivity, a necessary condition
(which may not be sufficient) is that some scattering mechanisms can
take place, so that the transmission, hence the thermal flux, is reduced.
Several scattering mechanisms exist in actual materials,
which may be intrinsic, as for inelastic anharmonic phonon-phonon scattering; 
or extrinsic, as for
elastic phonon scattering with defects. 
Experimental results showed that CNTs of
lengths $2.76~\mu$m may exhibit nearly ballistic
transport.\cite{YSYLM05} This justifies
that anharmonic scattering may be neglected if the elastic scattering
processes with defects are important.
The most simple defect that can be experimentally controlled is
isotope disorder, which amounts
here to replacing the usual $^{12}$C atom by one its $^{13}$C isotope. 
CNTs with isotope disorder have already been
synthetized\cite{SKPKZKSA05} and experimental results on 
boron nitride tubes\cite{CFAOIG06}
showed that isotope disorder could lead to dramatic changes in the
thermal conductivity. 
Moreover, isotopically disordered harmonic systems are also the simplest systems
that can be treated exactly with quantum statistics.

There are several methods to compute the thermal conductance
of isotopically disordered harmonic systems using quantum statistics
(we therefore disregard molecular dynamics techniques which give only
results within the classical framework).
These systems exhibit Anderson localization,\cite{MI70} which arises
from an interference effect of different scattering paths, and is thus
a manifestation of the ondulatory nature of the phonon vibrations.
The Green's function technique is then a very appealing method to compute the
thermal conductance in those systems, since it 
gives the exact transmission, 
has a computational cost scaling linearly with the system length, 
and is straightforward to parallelize since transport is
coherent. This allows to reach systems of lengths up to $L=2.5~\mu$m
(with up to 400,000 atoms). 
A Boltzmann approach, which describes the evolution of
the phonon distribution and therefore treats phonons as
particles and not as waves, gives only an average
description of the phonon flows in the system.
It cannot account for the Anderson localization of states, and it 
is therefore unclear
whether it can provide a good approximation to the exact transmission. 
A Boltzmann treatment of the transmission is however very interesting
since the method is computationally less expensive (no averages over
the disorder realizations have to be taken, the computational 
scaling with respect to the CNT index is more favorable, and the inclusion of
anharmonic scattering processes is easier than for Green's functions). 

In this paper, we will address three problems, 
systematically studying the length dependence of the
thermal conductivity with Green's function techniques:
\begin{enumerate}[(i)] 
\item does
  a well-defined conductivity (independent of the length) 
  exist for harmonic disordered CNTs, 
  or is the thermal transport anomalous? The theoretical results on
  heat transport in 1D chains suggest that the transport is anomalous,
  but can the asymptotic divergence exponent be estimated, or should
  incredibly long tubes be considered before such a regime is found?
\item how much is the
  thermal conductance reduced by isotopic disorder? How does this
  decrease depend on the temperature?
\item in order to reach even larger systems (larger diameters and/or
  lengths), a third questions is 
  whether the Boltzmann approach to thermal transport a good
  approximation for CNTs \emph{of experimental lengths}?
\end{enumerate}

The paper is organized as follows. We present briefly the notations and the 
numerical method in Section~\ref{sec:formalism} (see
Appendices~\ref{sec:Green_num} and \ref{sec:app_Boltz_num} for more precisions). 
We then turn to isotopically disordered one
dimensional chains in Section~\ref{sec:one_dimensional_results}, in
order to test the validity of the Boltzmann approach in this simple
case. Section~\ref{sec:flat_CNT_results} presents our new results on the
systematic study of the length dependence of the thermal conductivity
for CNTs. A conclusion is presented in Section~\ref{sec:conclusion}.

%
%

\section{Heat transport in (quasi) one-dimensional systems: A brief review}
\label{sec:heat_transport_1D}

\subsection{General features of heat transport in one-dimensional systems}

The theoretical and numerical results of heat transport in 1D chains
are not always acknowledged in computational studies of 
thermal conductivities of CNTs.
There have been many studies on the (non) validity
of Fourier's law in one dimensional chains. There are 
two important review papers on this topic,
written either from a mathematical\cite{BLRB00} or a physical\cite{LLP03} 
viewpoint. The main findings to this date may be summarized as follows:
\begin{enumerate}[(i)]
\item there are
  chains in which Fourier's law can be shown to 
  hold, in the sense that the conductivity does not depend on the
  system size. These situations arise 
  when translational symmetry is broken (through the addition of 
  some on-site potential in an anharmonic system for instance, 
  or when momentum is not conserved by the
  dynamics, see the discussion in Ref.~\onlinecite{BBO06} for more precisions),
  or for specific models
  such as systems with self-consistent stochastic reservoirs at each
  lattice sites\cite{BRV70} (a model recently
  studied mathematically in Ref.~\onlinecite{BLL04}); 
\item there are chains with anomalous conductivities, 
  {\it i.e} depending on the system size $L$. 
  If the interaction between the particles on the chain
  depends only on their relative distances (and no on-site potential is
  considered, in particular), then anomalous heat conduction is
  generally observed ($J/\nabla T  \sim L^\alpha$ 
  diverges as some power-law), 
  even with anharmonic interactions and/or mass disorder (see the
  references in Ref.~\onlinecite{LLP03}). The precise exponent for
  systems with anharmonic interactions is still a matter of debate, 
  with theoretical results\cite{NS02,LLP03b} ranging from
  $\alpha = 1/3$ to $\alpha = 2/5$, the most recent numerical results for
  Fermi-Pasta-Ulam chains giving values closer to $1/3$ 
  (see Ref.~\onlinecite{MDN07});
\item in some models, the scaling of the conductivity depends on
  the details of the thermal reservoirs, the way they are coupled to the
  system, or the boundary conditions used.\cite{LLP03,Dhar01}
\end{enumerate}
Most of these results are found from numerical (molecular dynamics)
simulations, though sometimes it is possible to prove mathematically the 
(non) existence of an $L$-independent conductivity for the model at hand.

More precise results on 
heat transport in isotopically disordered harmonic lattices (which is
the focus here) are recalled in 
Section~\ref{sec:theoretical_heat_harmonic_isotopic} when presenting
numerical results for 1D chains.


\subsection{Heat transport in CNTs}
\label{sec:heat_CNT}

The most popular method to compute conductivities is
nowadays molecular dynamics (MD). A noticeable exception is the Green's
function approach to defect scattering
of Refs.~\onlinecite{YW06} and~\onlinecite{YNW07}. Sometimes, the Boltzmann equation may be used as
well. We now give a brief review of MD studies.

Within the MD framework, 
the most popular approaches for thermal transport are (i) equilibrium methods, which rely on the Green-Kubo
formula; and (ii) nonequilibrium steady state (NESS) methods, where a
non-zero heat flux is created in the system, either through boundary forcing
(temperature difference) or using some mechanical forcing.\cite{Evans82}
Although the numerical procedures are by now standard computational
methods in molecular dynamics, they are not free from the general
limitations and issues concerning the convergence of the simulation
results. For instance, it is known that the numerical computation of
the conductivity relying on the Green-Kubo formula requires very long
time integrals of the heat current autocorrelation function, and averages
over many trajectories starting from different initial conditions. 
Besides, as the system size increases, the
typical correlation time increases as well.\cite{YWLL05}
In general, we expect a computational cost scaling as 
$\text{O}(L^2)$ since one force iteration of the MD solver requires
at least $\OO(L)$ operations, and the typical time (be it a correlation
time in the Green-Kubo case, or the relaxation time to reach a steady state for
NESS computations) is expected to scale as $\OO(L)$ as well. Moreover,
to obtain the temperature dependence of the thermal conductivity, many
simulations should be performed, each one at a different temperature.

Critical reviews of MD results have been written.\cite{MB05,LZ07} 
In particular, the upper limit to the
conductance given by the thermal ballistic conductance can be 
computed, and results of MD simulations compared to this
upper limit.\cite{MB05}
It is important to notice that MD results cannot be above the
ballistic upper limit to the conductance. When this is the case, 
it is likely that MD results are not converged.  
Besides, Ref.~\onlinecite{LZ07} showed that in some cases, the
simulation results depend on the boundary conditions and on the
simulation technique used.

In view of the MD computational scalings, only short nanotubes 
are computed (currently, the computations are usually restricted to 
system sizes $L \sim 0.2$~$\mu$m, whereas experimental lengths are of
the order $L \sim 2$~$\mu$m -- however, some MD results 
for $3.2~\mu$m CNTs are reported\cite{MITS06}) and only few 
works studied the size dependence of the thermal conductivity
for defect free CNTs in the framework of classical 
MD.\cite{Maruyama02,ZL05,LZ07,PXZ07,YWLL05,MITS06} This amounts to
considering only the effect of anharmonicity on the thermal conductance.
The results, in agreement with the predictions of the one-dimensional heat
transport theory, confirm the divergence of the conductivity with
increasing system sizes.
More recently, the effect of defects and impurities on the thermal
conductivity have been investigated. This includes isotopic 
disorder\cite{ZL05,SM06,BCYWC06,MITS06} and
functionalization.\cite{SBXOK04} In those cases, no systematic study
of length dependence was performed.

%
%

\section{Models for thermal transport}
\label{sec:formalism}

We present in this section two important models of transport: (i) the
Green's function approach, which solves the model in its full
atomistic details, and is exact for harmonic systems; 
(ii) the computationally less
expensive but more approximate Boltzmann treatment.
Both approaches will be used here to compute the decrease of the thermal
conductance arising from a disordered region embedded in an otherwise
perfect medium. We remark that the Green's function treatment
gives the exact transmission for a {\it given realization of the mass
  disorder}. Therefore, averages over the disorder realizations have
to be performed, unless the transmission function does not vary much
from one disorder realization to the other.
On the contrary, the Boltzmann approach already
gives some average transmission. 


\subsection{The Green's function treatment of thermal transport}

\subsubsection{Description of layered one-dimensional systems}
\label{sec:desription_layered_general}

Coherent thermal transport, presented in a pedagogical fashion 
in Ref.~\onlinecite{ZFM07}, involves expressions very similar to 
the usual expressions for electronic
transport.\cite{Datta2006,Datta2000}
The reason why Green's functions are introduced can be understood
by explicitly integrating the equations of motions
of the Heisenberg operators, projecting out the reservoir degrees of
freedom, and turning to Fourier variables since
only the steady state of the system is of interest. Such a derivation
follows the lines of the quantum Langevin equation,\cite{FKM65} and
the relationship with the nonequilibrium Green's function (NEGF) formalism
has been precised in Ref.~\onlinecite{DR06}. 
Rigorous proofs of the existence of
the NESS can be performed within the $C^\star$ scattering
formalism.\cite{AJPP07}

\begin{figure}
\includegraphics[width=7.3cm]{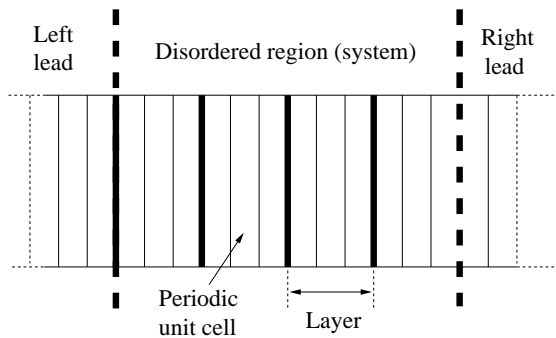}
\caption{ \label{fig:system}
  View of a layered 1D system. The infinite 
  system is decomposed into three regions: a semi-infinite left lead, a
  semi-infinite right lead (which are both assumed to be perfect), and
  a finite central disordered region. Assuming here that the atoms in
  the periodic unit cell interact with atoms located in the three
  neighboring cells on each side, a fundamental ``supercell'', called
  a layer, can be
  considered. Atoms in such a layer interact only with atoms in
  the two adjacent layers.
}
\end{figure}

The lattice thermal conductance arises from phonons,
\cite{AshcroftMermin,BdGdC01} which are
quantized displacements of 
harmonic lattices. 
We consider an infinitely extended system as depicted in 
Fig.~\ref{fig:system}. The Hamiltonian is
\[
H({\bf q},{\bf p}) = \frac12  {\bf q}^t K {\bf q} + \frac12 {\bf p}^t
M^{-1} {\bf p},
\]
where the infinite dimensional vectors 
\[
{\bf q} = (\dots,{\bf q}_{i,1},\dots,{\bf q}_{i,N},{\bf q}_{i+1,1},\dots)^t, 
\]
\[
{\bf p} = (\dots,{\bf p}_{i,1},\dots,{\bf p}_{i,N},{\bf p}_{i+1,1},\dots)^t
\]
stand respectively for displacements and momenta. 
The first index refers to the cell to which the atom
belongs, and the second indexes the atom within a cell. The cell in
question can be the periodic unit cell used to generate the system, but it
may also be some convenient supercell (see below).
The infinite dimensional matrix $M$ is the (diagonal) mass matrix of
the system:
\[
M = \left ( \begin{array}{ccc}
\ddots & & 0\\
& m_i & \\
0& & \ddots \\
\end{array} \right ).
\]
The matrix $K$ is the interatomic force constant matrix. It is assumed
to be short ranged in the sequel, so that an atom interacts only with
atoms located in a few neighboring unit cells.
Changing coordinates to mass weighted coordinates, 
the transport properties of the system can 
in fact be completely characterized by the harmonic matrix
\begin{equation}
  \label{eq:harmonic_matrix}
  A = M^{-1/2} K M^{-1/2}.
\end{equation} 

We restrict ourselves in this study to a disordered region embedded in
a perfect medium (see Fig.~\ref{fig:system}). 
The case of mass disorder is then dealt with by considering the mass of the
particles to be randomly distributed. In
the most physical case, namely isotopic disorder, 
the probability to have the mass $m$ at a given site is $1-c$, and the
probability to have a mass $m + \Delta m$ is $c$, where
$0 \leq c \leq 1$ denotes the disorder concentration. 
The assumptions on the system imply that the mass matrix is of the
general form
\begin{equation}
  \label{eq:mass_matrix_disordered}
  M = \left ( \begin{array}{ccc}
    M_L & 0 & 0 \\
    0 & M_\text{sys} & 0 \\
    0 & 0 & M_R \\
  \end{array} \right ),
\end{equation}
where in $M_L$ and $M_R$ all the masses are equal to $m$, while in 
$M_\text{sys}$ they are randomly distributed.

Carbon nanotubes are quasi one dimensional systems, that is, 
systems of finite size in the transverse directions and (infinitely) 
extended in the remaining direction.
We consider a fundamental 
supercell (a layer) composed of possibly several unit cells. 
The number of cells in this fundamental structure is determined by the
condition that an atom in a layer interacts only with atoms in the two
adjacent layers. The harmonic matrix therefore has the generic 
block tridiagonal shape
\begin{equation}
\label{eq:general_phonon_matrix}
A = \left ( \begin{array}{ccc}
A_L & \mathfrak{T}_L & 0 \\
\mathfrak{T}_L^t & A_\text{sys} & \mathfrak{T}_R \\
0 & \mathfrak{T}_R^t & A_R \\
\end{array} \right ).
\end{equation}
The (infinite) subsystems associated with the semi-infinite matrices
$A_L$, $A_R$ represent some reservoirs to
which the (finite) system $A_\text{sys}$ is coupled through the
interaction terms $\mathfrak{T}_L,\mathfrak{T}_R$. The block tridiagonal structure 
can be read off the expressions of the matrices
(for simplicity, we assume that 
there is no mass disorder in the first and the last layer of the
central region):
\[
A_L = \left ( \begin{array}{ccccc}
\ddots & \ddots & \ddots & 0 & \vdots\\
\ddots & \tau^t & a & \tau & 0 \\
& 0 & \tau^t & a & \tau \\
& \cdots & 0 & \tau^t & a \\
\end{array} \right ),
\]
\[
A_R = \left ( \begin{array}{ccccc}
a & \tau & 0 & \cdots & \\
\tau^t & a & \tau & 0 & \\
0 & \tau^t & a & \tau & \ddots \\
\vdots & 0 & \ddots & \ddots & \ddots \\
\end{array} \right ),
\]
\[
\mathfrak{T}_L = \left ( \begin{array}{ccc}
\vdots & \vdots &  \\
0 & 0 & \cdots \\
\tau & 0 & \cdots \\
\end{array} \right ), \qquad 
\mathfrak{T}_R = \left ( \begin{array}{ccc}
\vdots & \vdots & \\
0 & 0 & \cdots \\
\tau & 0 & \cdots \\
\end{array} \right ),
\]
while 
\[
A_\text{sys} = \left ( \begin{array}{cccccc}
a & \tau_1 & 0 & \cdots & & \\
\tau_1^t & a_1 & \tau_2 & 0 & &\\
0 & \tau_2^t & a_2 & \tau_3 & \ddots &\\
\vdots & 0 & \ddots & \ddots & \ddots & 0\\
& \dots & 0 & \tau_{N_\text{layers}-2}^t & a_{N_\text{layers}-2} & \tau_{N_\text{layers}-1}\\
& & \dots & 0 & \tau_{N_\text{layers}-1}^t & a \\
\end{array} \right ).
\]
In the above expressions, $a$ and $\tau$ are $3N_\text{at} \times
3N_\text{at}$ real matrices, $N_\text{at}$ denoting the number of atoms
in a layer. 
The matrix $a$ represents the interactions within a
layer, and the matrices $\tau^t, \tau$ model the interactions with
respectively the left and right neighboring layers
(For a matrix $\mathcal{M}$, $\mathcal{M}^t$ denotes the transpose
matrix, while $\mathcal{M}^\dagger$ denotes the
hermitian conjugate of $\mathcal{M}$ in the sequel).
If $N_\text{layers}$ is the number of layers composing the
disordered region, there are $N_\text{at} N_\text{layers}$ atoms in the
central part, and 
the matrix $A_\text{sys}$ is of size $3N_\text{at} N_\text{layers} \times
3N_\text{at} N_\text{layers}$. 
The matrices with underscripts $a_i, \tau_i, \tau_i^t$ differ from
the matrices $a, \tau, \tau^t$ because of the mass disorder 
(see Eqs.~(\ref{eq:harmonic_matrix})-(\ref{eq:mass_matrix_disordered})).

\subsubsection{Heat current in terms of Green's functions}

The Green's function of the whole system is defined as the limit
\[
G^+(\omega) = \lim_{\eta \to 0} \ (\omega^2+\ri \eta - A)^{-1}
\]
when this limit exists.\cite{Jaksic}
In numerical computations, 
the parameter $\eta$ is a small positive number. Knowledge of the 
Green's function implies the knowledge of the eigenvalues and density
of states of the system.
An interesting feature of the Green's function formalism in the
harmonic case is that 
the reservoir degrees of freedom can actually be 
projected out thanks to the linearity of the interactions, so that 
some effective dynamics only in the region of interest is
recovered. The effective Green's function for the central region is
\begin{equation}
  \label{eq:green_self_energy}
  G_\text{sys}^+(\omega) = \lim_{\eta \to 0}
  \ \left ( \omega^2+\ri\eta-A_\text{sys}-\Sigma^+_L(\omega)-\Sigma^+_R(\omega) \right )^{-1},
\end{equation}
where 
\begin{eqnarray}
  \label{eq:definition_self_energy}
  \Sigma^+_L(\omega) & = & \lim_{\eta \to 0}
  \ \mathfrak{T}_L^t (\omega^2+\ri\eta-A_L)^{-1} \mathfrak{T}_L, 
  \nonumber \\
  \Sigma^+_R(\omega) & = & \lim_{\eta \to 0} 
  \ \mathfrak{T}_R (\omega^2+\ri\eta-A_R)^{-1} \mathfrak{T}^t_R.
\end{eqnarray}
The operators $\Sigma_L^+,\Sigma_R^+$ are self-energies, which 
model the coupling of the system with the contacts. In particular, the
imaginary part of the self energy is usually associated with some
lifetime (due to phonons flowing out of the central region). 
This is maybe more easily understood in the quantum Langevin
framework, where the self energy is the Fourier transform of the
friction kernel.\cite{DR06}

When there are no incoherent scattering processes, 
the current is given as the superposition of phonons going from the
left reservoir to the right one, minus the flow of phonons going from
the right reservoir to the left one. 
Therefore, phonons entering the system can be traced back to the reservoir
where they come from and so, the summation is
restricted to phonons with positive group velocity for phonons coming
out of the left reservoir, and phonons with negative
velocities for the ones coming out of the right reservoir. 
When the 
reservoirs are at different temperatures (respectively, $T_L$
and $T_R$), the heat current flowing through the system 
is\cite{Datta2006,ZFM07,DR06} 
\begin{equation}
  \label{eq:heat_current}
  J(T_L,T_R) = \int_{0}^{+\infty} \frac{\hbar \omega}{2\pi} \cT(\omega) 
  (f_{T_L}(\omega)-f_{T_R}(\omega)) \, d\omega,
\end{equation}
where the transmission factor $\cT(\omega)$ is 
\begin{equation}
  \label{eq:transmission}
  \cT(\omega) = \tr \left [ \Gamma^+_L(\omega) G_\textrm{sys}^+(\omega) \Gamma^+_R(\omega)
    \left ( G_\textrm{sys}^+(\omega) \right )^\dagger \right ]. 
\end{equation}
In the above expressions, 
\[
\Gamma^+_\alpha(\omega) = -2 \, \text{Im}\left(
\Sigma^+_\alpha(\omega) \right) \ (\alpha = L,R)
\]
is related to the imaginary part of the self-energies
and the functions $f_T$ are the Bose-Einstein distributions
\[
f_T(\omega) = \left (\exp\left(\frac{\hbar \omega}{k_\text{B}
  T}\right)-1 \right )^{-1}.
\]
The expression of the thermal conductance is exact since the system is harmonic.
The practical computation of the transmission is
presented in Appendix~\ref{sec:Green_num}.

Introducing $t(\omega) = \Gamma_L(\omega)^{1/2}
G_\textrm{sys}^+(\omega) \Gamma_R(\omega)^{1/2}$, the transmission can be
rewritten as $\mathcal{T}(\omega) = \tr (t(\omega)t(\omega)^\dagger)$. It
is therefore nonnegative. In fact, when there is no disorder, the
transmission is ballistic and equal to the number of conducting
channels at this pulsation. In the presence of mass disorder, the
transmission is lower than the ballistic transmission.

\subsubsection{Thermal conductance and thermal conductivity}

The thermal conductance associated with the heat current~(\ref{eq:heat_current})
is obtained in the limit $\Delta T
= T_R - T_L \to 0$ as
\begin{eqnarray}
  \label{eq:thermal_conductance}
  g(T) & = & \lim_{\Delta T \to 0} 
  \frac{\dps J\left(T+\frac{\Delta T}{2},T-\frac{\Delta
  T}{2}\right)}{\Delta T} \nonumber \\
  & = & 
  \int_{0}^{+\infty} \frac{\hbar \omega}{2\pi} \cT(\omega) 
  \frac{\partial f_{T}(\omega)}{\partial T} \, d\omega.
\end{eqnarray}
A remarkable fact about the heat current is that the associated
thermal conductance is quantized,\cite{RK98} the quanta of thermal
conductance being
\begin{equation}
  \label{eq:QuantaThermalConductance}
  g_0(T) = \frac{\pi^2 k_\text{B}^2 T}{3h}.
\end{equation}
This expression is obtained in the limit $T \to 0$ for a single
acoustic branch in the ballistic regime.
The quantum of thermal conductance has been experimentally 
measured.\cite{SHWR00} In the sequel, conductances will most often be
normalized with respect to the quantum of thermal conductance:
\begin{equation}
  \label{eq:normalized_thermal_conductance}
  \overline{g}(T) = \frac{g(T)}{g_0(T)}.
\end{equation}
The dimensionless quantity $\overline{g}(T)$ is called
``normalized conductance'' in the results presented below.

For 1D systems, the thermal conductivity $\kappa$ 
is the thermal conductance divided by the
system cross section $\mathcal{A}$, and multiplied by the system
length $L$:
\begin{equation}
  \label{eq:conductivity}
  \kappa = \frac{g L}{\mathcal{A}}.
\end{equation}
The notion of cross section depends on the
system considered and on the conventions used (see 
Section~\ref{sec:CNT_model} for our conventions in the case of
CNTs). From this definition, it is however clear that the existence of a
well-defined (length independent) non-vanishing thermal conductivity requires a
decrease of the thermal current (or thermal conductance) as $1/L$. 
Since the thermal conductivity is not well-defined {\it a priori}, the
thermal conductance is a much better notion to handle.

The practical computation of the thermal conductance 
using~(\ref{eq:thermal_conductance}) therefore boils down to
computing the transmission factor $\cT(\omega)$ given 
by~(\ref{eq:transmission}) for the whole
phonon spectrum (see Appendix~\ref{sec:Green_num}).


\subsection{The Boltzmann approximation of coherent transport}

Boltzmann method is a less expensive but more approximate method than
the Green's function treatment for the computation of the thermal
conductance.
It allows to compute conductivities for very long systems, 
and therefore to study the length dependence of the conductivity. For
instance, the effect of anharmonicity (in terms of 3 and 4 phonon
scattering processes) has been considered for tubes of
lengths up to $L=1$~m.\cite{MB05_bis} 
Recent derivations of the Boltzmann equation from microscopic dynamics
have been presented in Refs.~\onlinecite{Spohn06} and~\onlinecite{SL07} 
in the so-called kinetic limit (vanishingly small mass disorder or 
anharmonicities, long time limit, interatomic spacing going to zero). 
The proofs are sometimes only formal, the central object in the 
overall procedure being the Wigner distribution of phonons. 
The Boltzmann equation therefore describes some average behavior of the phonons.

The central object describing this average behavior is the phonon
distribution $n_j(\omega,x,t)$. The index $j$ labels the phonon branch, 
$\omega$ is the phonon pulsation, while $x$
is the position along the system. The quantity $n_j(\omega,x,t)$ therefore
counts the number of phonons of pulsation $\omega$ 
in the $j$-th branch at a point $x$ at time $t$. 
The evolution of the phonon distribution is governed 
by the Boltzmann equation, which consists of two terms: 
\begin{itemize}
\item a transport term, which models the free 
  flow of phonons through the harmonic system of reference;
\item a collision term, which models the rate at which phonons from
  one branch are scattered into another branch due to the defects and/or
  impurities in the actual system. 
\end{itemize}
Since we do not consider anharmonic effects in this study but only
mass defects,
the energy is conserved in the collision processes 
and the only scattering mechanism is the scattering from a phonon in a
given branch to a phonon in another branch {\it at the same energy}.  

\subsubsection{The Boltzmann equation}

For the system introduced in
Section~\ref{sec:desription_layered_general}, 
a phonon at a wavevector $k$ is
an eigenvector of the dynamical matrix of the perfect system:
\[
D(k) = a + \tau \text{e}^{-\ri k l_0} 
+ \tau^t \text{e}^{\ri k l_0},
\]
where $l_0$ is the lattice parameter (determined by the length of
the periodic unit cell) and
$k$ belongs to the first Brillouin zone $[-\pi/l_0,\pi/l_0]$.
More precisely, the $j$-th phonon is 
$U_j = (\text{e}^{-\ri k n l_0} u_j)_{n \in \mathbb{Z}}$ 
where $u_j$ is such that
\[
D(k) \, u_j = \omega_j(k)^2 \, u_j.
\]
Notice that since $a$ and $\tau$ are real matrices and $\omega_j$ is a
real number, $U_j^\star$ is also a
phonon since it is an eigenvector of the dynamical matrix $D(-k)$ for the pulsation
$\omega_j(-k) = \omega_j(k)$.
For a given pulsation $\overline{\omega} \geq 0$, 
the associated phononic branches $j$ at this pulsation are
all the solutions of the equation
\begin{equation}
  \label{eq:condition_decay}
  \omega_j^2(k) = \omega_j^2(-k) = \overline{\omega}^2, \quad 0 \leq j
  \leq 3 N_\text{at},
\end{equation}
for some $k$ in the Brillouin zone. 
We denote by $N \equiv N(\overline{\omega})$ the number of branches at this
pulsation, and by $n_{j,\sigma}(\omega,x,t)$ the density of phonons
of the $j$-th branch ($j=1,\dots,N(\overline{\omega})$ 
upon reordering). 
The index $\sigma = \pm 1$ labels the
solutions $\omega_j^2(k)=\overline{\omega}^2$ depending on the sign of
the phononic velocity: $\sigma=+1$ corresponds
to $k$ points such that
\[
\omega_j(k_{j,+1}) = \overline{\omega}, \qquad v_j = \frac{\partial \omega_j}{\partial k}(k_{j,+1}) > 0,
\]
whereas the velocity is negative at points $k_{j,-1} = -k_{j,+1}$, and is actually
equal to $-v_j$. 

The Boltzmann equation reads
\begin{widetext}
\begin{equation}
  \label{eq:Boltzmann_n}
  \left ( \partial_t + \sigma \, v_j \partial_x \right ) n_{j,\sigma}(\omega,x,t)
  = \sum_{(j',\sigma') \not = (j,\sigma)}
  \frac{ W_{(j,\sigma),(j',\sigma')} }{v_j} \, n_{j',\sigma'}(\omega,x,t) 
  - \frac{ W_{(j',\sigma'),(j,\sigma)} }{v_{j'}} \, n_{j,\sigma}(\omega,x,t).   
\end{equation}
\end{widetext}
The left-hand side of the above equation is the transport part of the
Boltzmann equation.
Notice also that $\omega$ does not enter explicitly the above
equation. It can therefore be considered as a continuous parameter
indexing a set of independent equations. 

The matrix $W=[W_{(j,\sigma),(j',\sigma')}]_{1 \leq j,j' \leq N, \
  \sigma,\sigma' = \pm 1}$ describes the interactions between the
  branches, {\it i.e.} the collision term. It is such that 
for all $(j,\sigma) \not = (j',\sigma')$, 
\[
W_{(j,\sigma),(j',\sigma')} \geq 0, 
\]
\[
W_{(j,\sigma),(j',\sigma')} = W_{(j',\sigma'),(j,\sigma)}, 
\]
\[ 
W_{(j,-\sigma),(j',-\sigma')} = W_{(j',\sigma'),(j,\sigma)}.
\]
The above conditions have physical interpretations:
(i) the first condition (non-negativity of the matrix elements)
means that the scattering term models the decay of phonons of one branch 
into phonons of another branch; (ii) the second condition ensures the
conservation of the total phonon number; (iii) the last condition is some symmetry condition with
respect to the propagation direction of phonons.
The precise form of the scattering matrix depends on the problem at hand; see
below the expression~(\ref{eq:facteur_isotopic_disorder}) 
in the case of isotopically disordered lattices.

\subsubsection{The collision term}

The expression of the scattering matrix are derived from a perturbative
approach (based on the Fermi Golden Rule):\cite{Tamura83,WSC02,Niels}
\begin{equation}
  \label{eq:facteur_isotopic_disorder}
  W_{(j,\sigma),(j',\sigma')} = l_0 \, \frac{\text{Var}(m)}{\langle m \rangle^2}
  \, \overline{\omega}^2 \, \sum_{l = 1}^{N_\text{at}} \left |
  {\bf u}^\star_{j,\sigma}(l) \cdot {\bf u}_{j',\sigma'}(l) \right |^2,
\end{equation}
where $\langle m \rangle$ and 
$\text{Var}(m) = \langle m^2 \rangle - \langle m \rangle^2$ 
are the average mass and the variance
of the mass disorder respectively.
The three dimensional vector ${\bf u}_{j,\sigma}(l)$ is the part of $u_j$
corresponding to the three degrees of freedom of the $l$-th atom in
the unit cell for phonon displacements $u_j$ 
computed for a perfect lattice.

The scattering term~(\ref{eq:facteur_isotopic_disorder}) agrees with
the mathematical results obtained by the kinetic limit of the
microscopic dynamics in the case of a simple one-dimensional 
chain.\cite{SL07}
From a numerical viewpoint, the
expression~(\ref{eq:facteur_isotopic_disorder}) is very interesting
since it allows an analytic computation of the scattering rates once
the phonon spectrum has been computed. 
Appendix~\ref{sec:app_Boltz_num} presents the details of the numerical solution
of~(\ref{eq:Boltzmann_n}) using the scattering 
term~(\ref{eq:facteur_isotopic_disorder}).

\subsubsection{Transmission properties using the Boltzmann equation}

Thermal properties are computed for nonequilibrium steady states where
a heat current flows through the system. This is done by setting
appropriate boundary conditions, and waiting for the system to equilibrate.
The boundary conditions for~(\ref{eq:Boltzmann_n}) are suited to the
transmission of a single phonon from the left (hot) 
reservoir to the right (cold) one:
\[
n_{j,+}(\omega,0,t) = 1, \qquad n_{j,-}(\omega,L,t) = 0.
\]
The numbers $n_{j,+}(\omega,L,t)$ and $n_{j,-}(\omega,0,t)$ are respectively the
proportion of transmitted and reflected phonons. These proportions
are computed using the Boltzmann equation, and this defines the
transmission factor.

When a stationary regime is reached ($\partial_t n_{j,\sigma} = 0$ for
all $j,\sigma$), the phonon distributions do not depend on time
anymore, and we drop the variable $t$ in our notations.
It is also easily seen that the transmission coefficient is 
independent of the position $x$, so that
\begin{eqnarray}
T(\omega,x) & = & N(\omega) \frac{\dps \sum_{j=1}^{N(\omega)} 
  \left [ n_{j,+}(\omega,x)-n_{j,-}(\omega,x)\right]v_j(\omega)}
{\dps \sum_{j=1}^{N(\omega)} v_j(\omega)} \nonumber \\
& = & N(\omega) \frac{\dps \sum_{j=1}^{N(\omega)}  
  n_{j,+}(\omega,L) \, v_j(\omega)}{\dps \sum_{j=1}^{N(\omega)}  v_j(\omega)}.
\end{eqnarray}
This coefficient is the
Boltzmann equivalent of the transmission function $\cT(\omega)$
computed using Green's function techniques.

%
%

\section{One dimensional chains}
\label{sec:one_dimensional_results}

Perfect one-dimensional harmonic lattices with nearest neighbor
interactions are described by the Hamiltonian
\[
H = \sum_j \frac{p_j^2}{2 m} + \frac12 k (q_{j+1}-q_j)^2.
\]
This model fits in the general framework
presented in Section~\ref{sec:formalism} when taking $N_\text{at} = 1$
atoms per unit cell and layers of $N_\text{layers} = 1$ unit cell.
The intralayer interactions are then described by the $1 \times 1$
matrix $a = [2k/m]$, while the interlayer interactions $\tau =
\tau^t = [-k/m]$.

In this section, we will work in reduced units of masses and lengths, and 
consider a one dimensional
chain with unit lattice spacing and particles of masses 1. 
Isotopic
disorder is modelled by replacing with probability $0 < c < 1$ the
mass of a particle by $1 + \delta$ with $\delta = \Delta m/m$.

\subsection{Heat transport in isotopically disordered harmonic lattices}
\label{sec:theoretical_heat_harmonic_isotopic}

Heat transport in isotopically disordered harmonic lattices  
has been thoroughly studied.\cite{MI70,RG71,CL71,Ishii73,OcL74} 
In particular, 
the case a finite disordered chain connected to two semi infinite perfect
chains has been considered,\cite{RG71} and the corresponding 
theoretical results were rederived from a different perspective 
and extended in Refs.~\onlinecite{OcL74} and~\onlinecite{Dhar01}. 
It is suggested in Ref.~\onlinecite{RG71} that the thermal 
conductivity diverges as $\sqrt{L}$ ($L$ being the chain length).
This has been shown rigorously\cite{KPW78} for an
analogous continuum wave model (open boundary conditions 
and disorder). 

Heuristic arguments can be employed to back up the $\sqrt{L}$
divergence of the thermal conductivity. 
Denoting explicitly the length dependence of the 
transmission function by $\cT_L(\omega)$, the exact transmission 
is rewritten as
\begin{equation}
  \label{eq:transmission_gamma}
  \cT_L(\omega) \simeq \exp ( - L/l_\text{loc}(\omega)),
\end{equation}
since it can be shown that the limit
\[
\lim_{L \to +\infty} -\frac{1}{L} \ln \cT_L(\omega) = \frac{1}{l_\text{loc}(\omega)}
\]
exists for all $\omega > 0$. 
This result was proved in Ref.~\onlinecite{MI70} (but
earlier obtained in the limit of low concentration
of defects\cite{Rubin68}). The proof is based on a theorem by Furstenberg on 
the product of random matrices.\cite{Furstenberg63}
The behavior of $l_\text{loc}$ around $\omega^2 = 0$ can also be 
precised.\cite{MI70,OcL74} For
isotopic disorder, it can be shown that
\begin{equation}
  \label{eq:transmission_gamma_asymptotic}
  \lim_{\omega^2 \to 0} \frac{1}{\omega^2 \, l_\text{loc}(\omega)} 
  = \frac{\text{Var}(m)}{\langle m \rangle^2} 
  = \frac{c(1-c)}{4} \frac{\delta^2}{(1 + c\delta/2)^2}.
\end{equation}
This shows that high a frequency implies a small $l_\text{loc}$ and
therefore the associated eigenmodes are strongly
localized, so that only the low
frequency modes contribute to transport; in fact, only a fraction
$\text{O}(L^{-1/2})$ of them since
\[
\cT_L(\omega) \simeq \exp \left ( - \frac{\text{Var}(m)}{\langle m
  \rangle^2}  L \omega^2 \right )
\] 
when $\omega \to 0$.
This explains therefore the $L^{-1/2}$
decay of the thermal conductance.

\subsection{Comparison of the Green's function and Boltzmann treatments}

\begin{figure}
\center
\includegraphics[width=7.3cm]{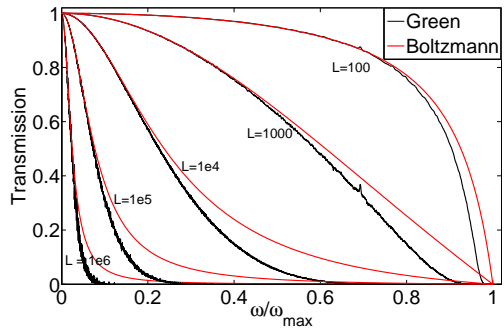}
\caption{\label{fig:trans_1D} (color online)
  Averaged transmission function $\cT_L(\omega)$ as a function of
  $\omega$, for different system sizes. 
  From top to down: increasing system sizes from $L=100$ to $L=10^6$.
  The black curves are the transmissions computed with a Green's
  function approach, and the red curves are obtained from the
  Boltzmann formula~(\ref{eq:Boltz_1D}).
}
\end{figure}

The transmission is computed using the Green's function approach 
at $N_\omega = 5000$ points uniformly spaced
in the range $[0,\omega_\text{max}]$, with $\eta/\omega_\text{max}^2 = 10^{-12}$ (the self energies having
an analytical expression, see for instance\cite{Economou}). The
averages are taken over $N_\text{disorder} = 100$ realizations of the isotopic disorder for
chains of length $L=10^6$, while $N_\text{disorder} = 1000$ for
$L=10^5$, and $N_\text{disorder} = 10^4$ for $L=10^2 - 10^4$. 
The average transmission functions computed from~(\ref{eq:transmission}) are presented in
Fig.~\ref{fig:trans_1D} in the
case $\delta = 1/12$ and $c=0.5$. The mass
variation $\delta$ is the mass variation corresponding to substituting $^{12}$C
with $^{13}$C. 

The transmission predicted by the Boltzmann approach
can be computed analytically in this simple case since there is only a single
branch and there are therefore only
two conducting states for a given pulsation (due to the symmetry $\omega(k) =
\omega(-k)$). It can be shown (see Appendix~\ref{sec:app_Boltz_1D}) that
\begin{equation}
  \label{eq:Boltz_1D}
  T_L(\omega) = \left( 1 + \frac{L}{l_\text{Boltz}(\omega)} \right )^{-1}, 
\end{equation}
with 
\[
\frac{1}{\omega^2 \, l_\text{Boltz}(\omega)} =  
\frac{\text{Var}(m)}{\langle m \rangle^2}
\left(1-\frac{\omega^2}{4}\right)^{-1}.
\]
Notice that this expression of the transmission agrees at second order 
in $\omega$ 
with~(\ref{eq:transmission_gamma}) in the limit $\omega \to 0$.

The comparison between the transmissions computed with the Green's
function and the Boltzmann approaches shows that the Boltzmann
treatment is a reasonable approximation for low frequency modes, 
but predicts a transmission which is too large for
higher frequencies. The critical frequency at which the Boltzmann
transmission starts to depart significantly from the Green's function
transmission decreases with the system size. 
Therefore, we expect the conductances to agree in
the low temperature regime for all system lengths, and the relative 
error to increase with the system length at larger temperatures where the
higher frequency part of the transmission are taken into account.

\begin{figure}
\center
\includegraphics[width=7.3cm]{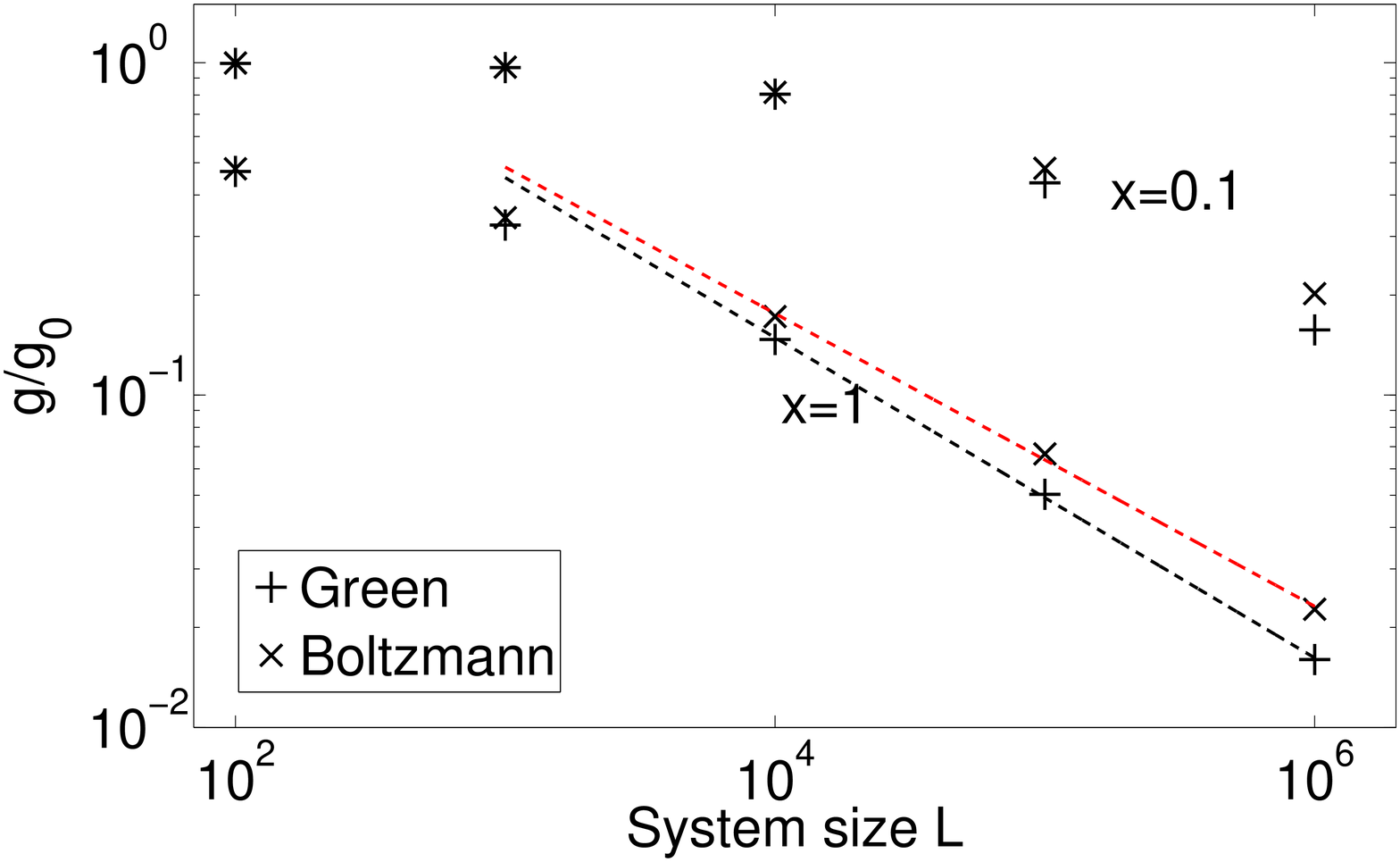}
\caption{\label{fig:conductance_1D} (color online)
  Conductances computed using the
  Green's function approach and a Boltzmann treatment
  for two values of $x=\hbar \omega/ k_\text{B} T$. The asymptotic regime where
  $\overline{g} \sim L^{-\alpha}$ is attained only for $x=1$, with
  $\alpha \simeq 0.48$ for the Green's function approach, and $\alpha
  \simeq 0.44$ in the Boltzmann case.
}
\end{figure}

This is indeed confirmed by the scalings of the normalized thermal conductance in 
Fig.~\ref{fig:conductance_1D}, for different values of 
$x = \hbar \omega/ k_\text{B} T$.
As expected, the asymptotic scaling
$L^{-1/2}$ for the thermal conductance is reached for systems long
enough when the transmission is computed using the Green's function
approach. This could be anticipated since the Green's function results
are exact for harmonic systems.
On the other hand, the conductances predicted by the Boltzmann
approach for systems long enough, 
are larger than the conductances obtained with the Green's
function approach. The Boltzmann treatment does 
not predict the right asymptotic scaling either, though the
discrepancy is not too large. This is due to
the fact that the Boltzmann transmission does not decay fast enough
with the system size for larger values of $\omega$.

\begin{figure}
\center
\includegraphics[width=7.3cm]{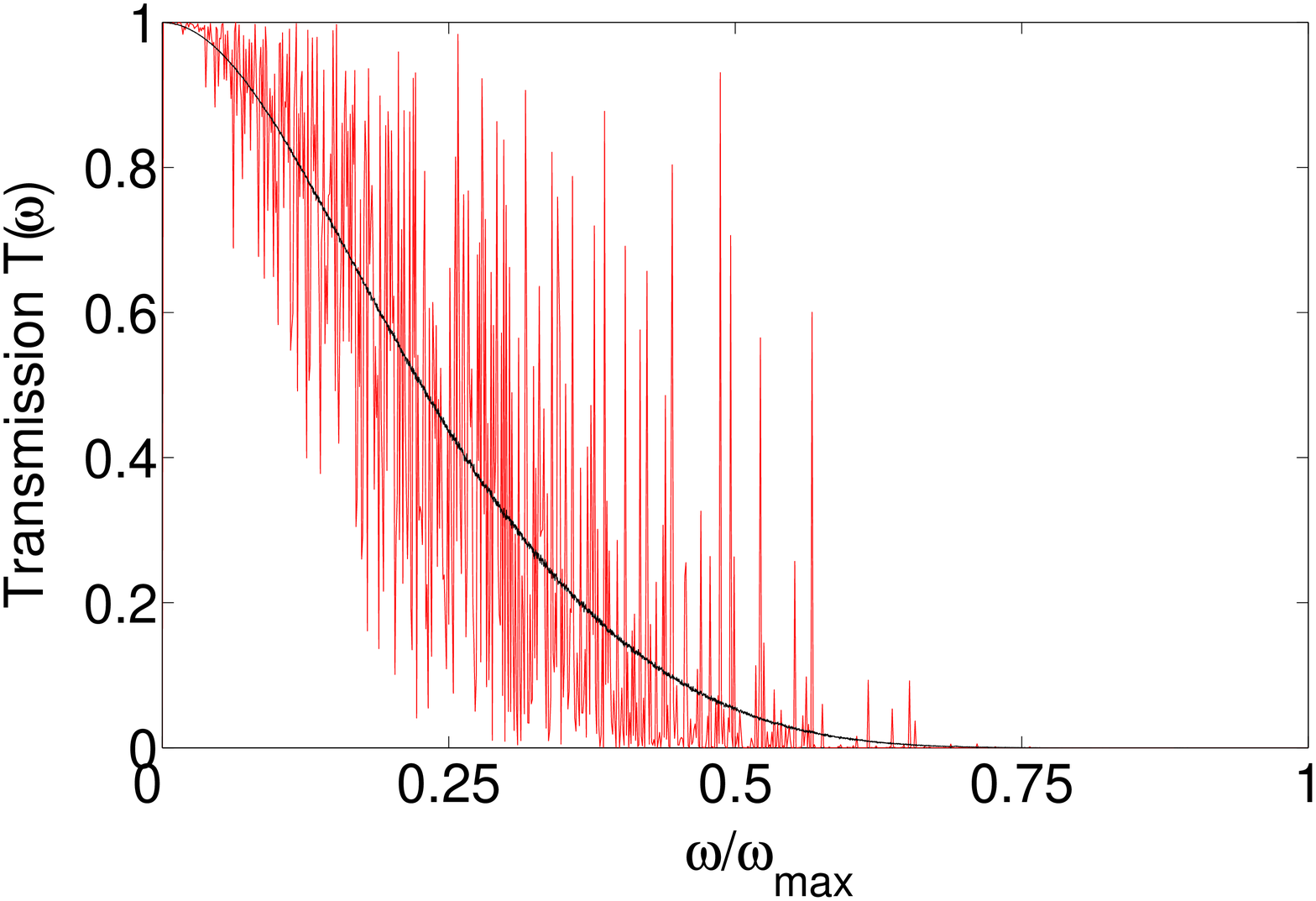}
\includegraphics[width=7.3cm]{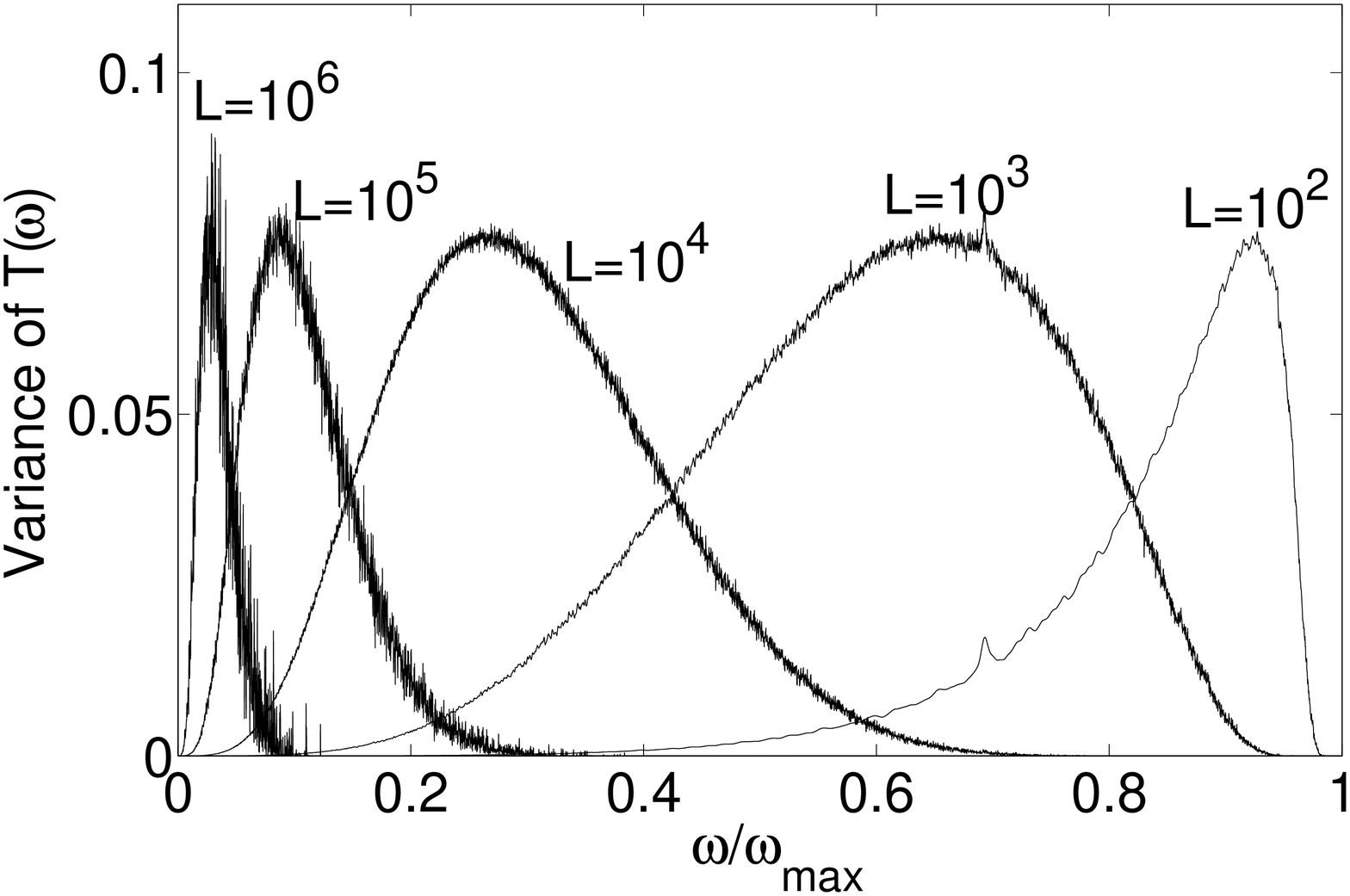}
\caption{\label{fig:single_trans_1D} (color online)
  Top: Transmission computed using the Green's function technique
  for a one dimensional chain of length $L=10^4$:
  average transmission (black) and transmission for a single
  realization of the mass disorder (red).
  Bottom: Variance of the transmission functions computed using the
  Green's function techniques as a function of the
  pulsation, for different system lengths.
}
\end{figure}

An interesting feature of the transmission function in the one
dimensional case is that 
the maximal variance of $\cT(\omega)$ over the
realizations of the transmission computed with different mass
disorders does not decrease with the increasing
size of the system (though the range of pulsations where the variance
is not small decreases).
However, the thermal conductance, which is the integral of the
transmission function, does not vary much from one realization of the
mass disorder to the other.
Fig.~\ref{fig:single_trans_1D} (Top) presents a plot of a 
the transmission function for a given realization of the mass
disorder, which is compared to the average transmission, while 
Fig.~\ref{fig:single_trans_1D} (Bottom) plots the variance of the
transmission as a function of the pulsation.
Therefore, it is not true that, when the size of the disordered region increases,
all the transmission functions look similar (no thermodynamic limit
seems to be reached). There remains some intrinsic variability, which
can be used to determine exponential localization lengths.\cite{ARTBL07}
This is also reminiscent of the fact that, in molecular dynamics studies of
harmonic disordered chains,  
non smooth steady temperature profiles are obtained when only 
a single realization of the disorder is considered.\cite{LVAY06}

%
%

\section{Numerical results for carbon nanotubes}
\label{sec:flat_CNT_results}


\subsection{Description of the model}
\label{sec:CNT_model}

We use two distinct models to simulate CNTs. 
With the first one, labelled as 'ab-initio' in the sequel, 
the tube is simulated using the interatomic force constants (IFCs)
computed from density functional theory calculations,\cite{BdGdC01}
using the PWCSF package of the QUANTUM-ESPRESSO distribution\cite{QuantumEspresso}
on a (5,5) CNT of experimental (curved) geometry.\footnote{
  The computations were performed using the local density approximation,\cite{LDA} 
  norm-conserving pseudo-potentials,\cite{pseudo,pseudo2} and a plane-wave expansion up to 
  55 Ry cut-off. Brillouin-zone sampling was performed on a 
  $1 \times 1 \times 32$ Monkhorst-Pack mesh, with a Fermi-Dirac smearing
  of 0.02~Ry. 
  The dynamical matrices are calculated on a $1 \times 1 \times 12$ grid of $q$-points,
  and Fourier interpolation is used to get the dynamical matrices on
  a finer mesh of $q$-points.
  The theoretical radius of the (5,5) CNT is 6.412~Bohrs.
}
With the second one, labelled as 'empirical' in the sequel, 
the tube is simulated with a flat geometry using the empirical 
force constants of graphene given in Ref.~\onlinecite{SDD98}.
The empirical model is used to simulate (5,5) and (10,10) CNTs.
In any cases, the models
are used consistently within both the Green's function and Boltzmann
approaches. 
We restrict ourselves to armchair nanotubes
for simplicity. Armchair nanotubes
are very important for applications since they are
metallic (here we compute only the phononic thermal
conductance). 
We expect the results presented below to be 
robust with respect to the system chirality, because 
the ballistic conductance does not change much with the chirality.\cite{MB05}

\begin{figure}
\center
\includegraphics[width=7.3cm]{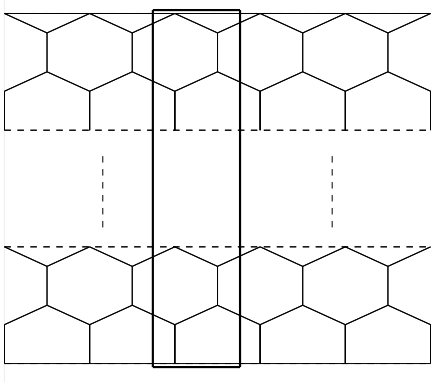}
\caption{\label{fig:unit_cell}  
  Armchair nanotube, with 
  interactions up the third neighbor. The unit cell of the periodic structure
  is represented by the box.
}
\end{figure}

In order to have a block tridiagonal form for the
interaction matrix within the ab-initio model, 
only interactions within a cut-off radius $R_{\rm cut} = 12$~Bohrs are
taken into account. 
In this case, it is necessary to consider a supercell 
composed of two elementary cells depicted in Fig.~\ref{fig:unit_cell} as a
unit building block.
The projection procedure of Ref.~\onlinecite{Mounet05} was used to ensure 
that acoustic sum rules are satisfied for the truncated IFC matrix. 

When the empirical model of Ref.~\onlinecite{SDD98} is used, 
carbon nanotubes are modeled as 
graphene nanoribbons with periodic boundary
conditions in the transverse direction, and interactions up the
third neighbors are taken into account. Due to these simplifying assumptions, 
the in-plane and out-of-plane motions can be computed
separately since their contributions are independent. 
In this case, the elementary cell depicted in Fig.~\ref{fig:unit_cell} can be chosen as a
unit cell. Although less precise (though, as will be shown below, many results are in very 
good agreeement with the results obtained with the {\it ab-initio} based computations), 
this model is less expensive and was therefore 
used to study CNTs of larger diameters. 

\begin{figure}
\center
\includegraphics[width=7.3cm]{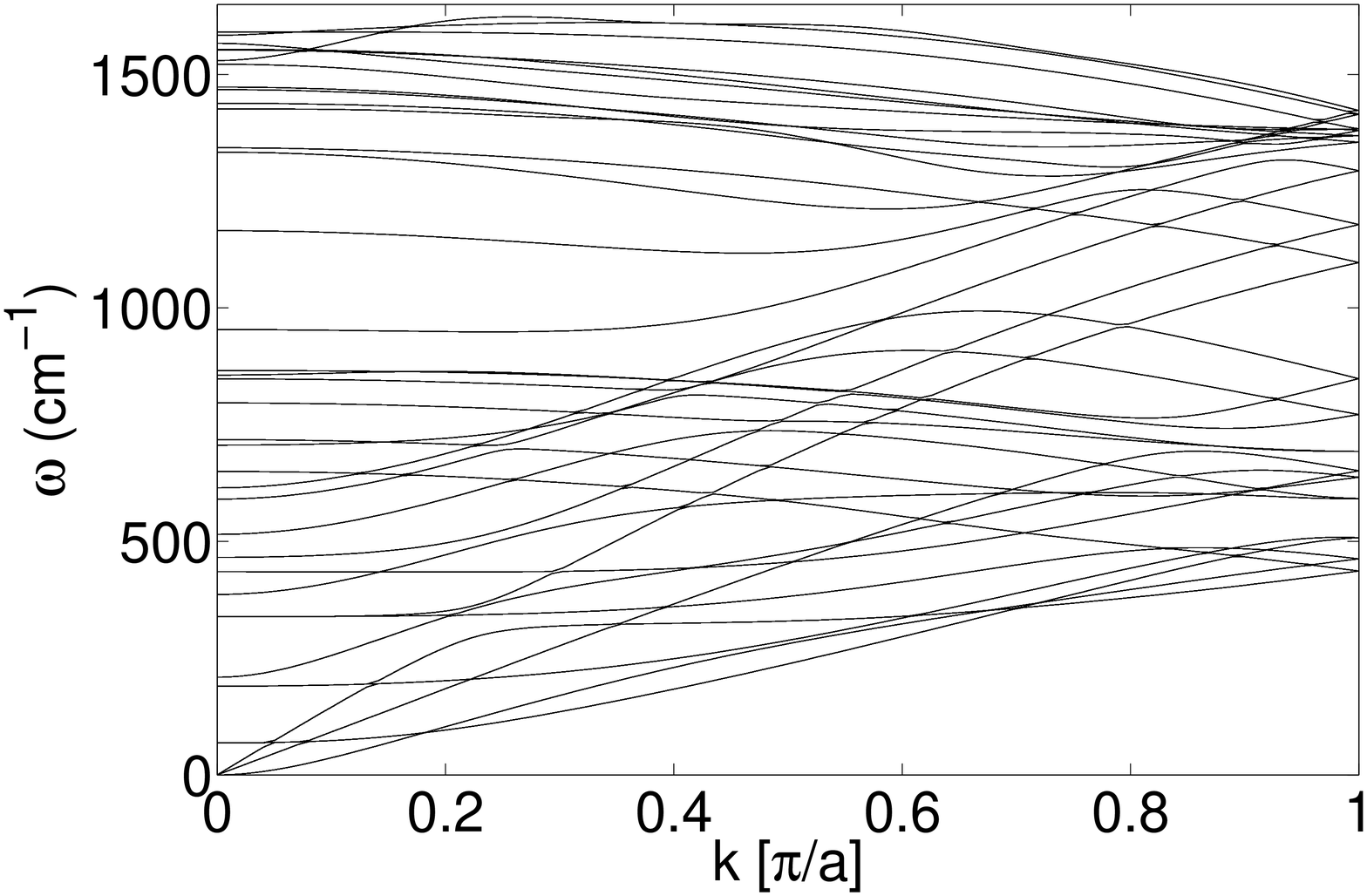}
\includegraphics[width=7.3cm]{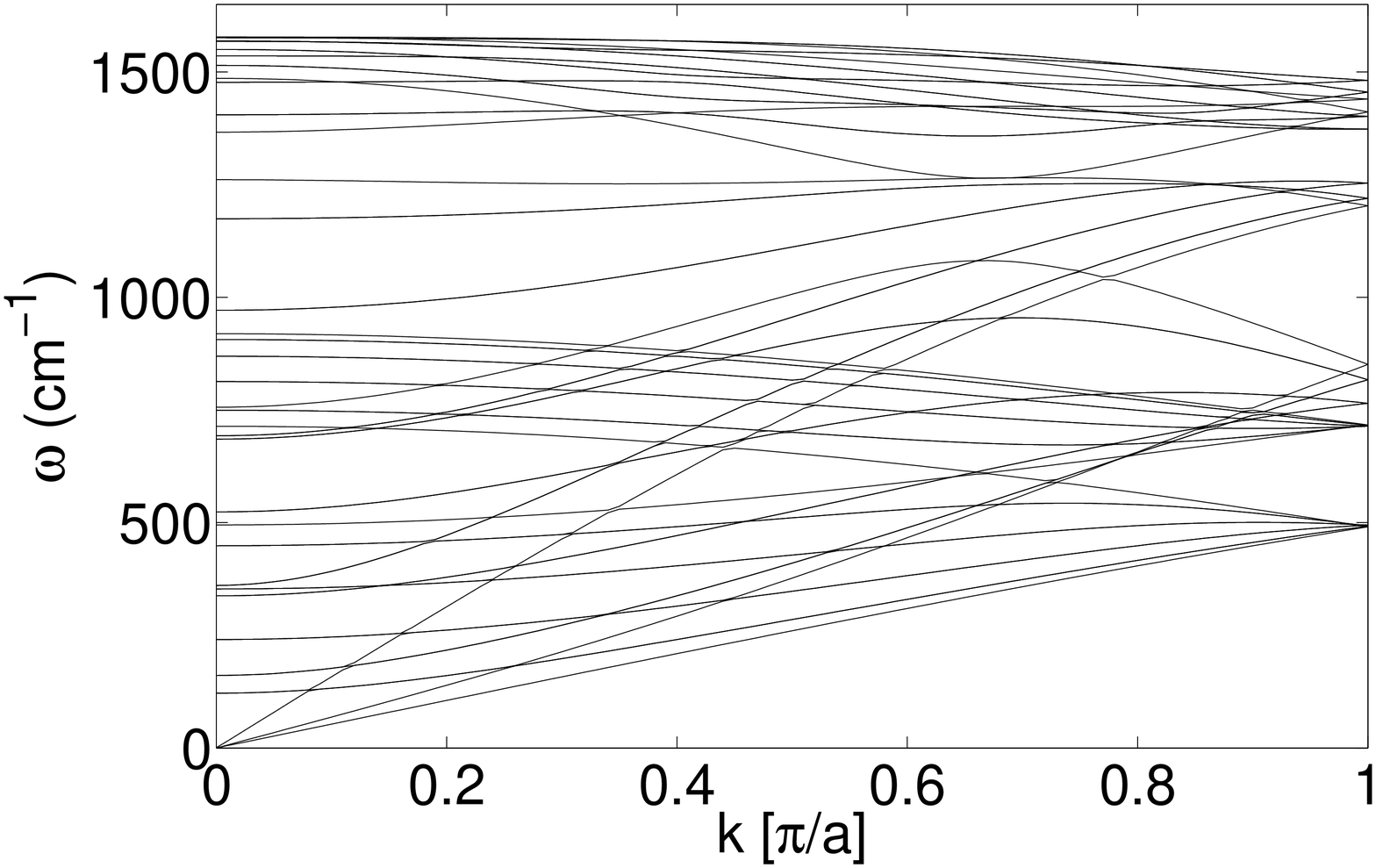}
\caption{\label{fig:SpectrumFlatCNT} 
  Phonon spectrum of a (5,5) CNT. Top: {\it Ab-initio} computed
  interatomic forces. Bottom: Empirical model described in 
  Section~\ref{sec:CNT_model}.
}
\end{figure}

We compare in Fig.~\ref{fig:SpectrumFlatCNT} phonon dispersions
for (5,5) armchair tubes obtained with the empirical model and 
with {\it ab-initio} computed IFC.
The agreement is fair enough, though the number of acoustic modes is
not correct, and the lowest modes do not have a quadratic dispersion
relation as is the case for real nanotubes. 
Despite that, the thermal transport results for both models should agree quantitatively 
as shown by a comparison of ballistic conductances computed with the empirical model
used here and {\it ab-initio} reference results.

\begin{figure}
\center
\includegraphics[width=7.3cm]{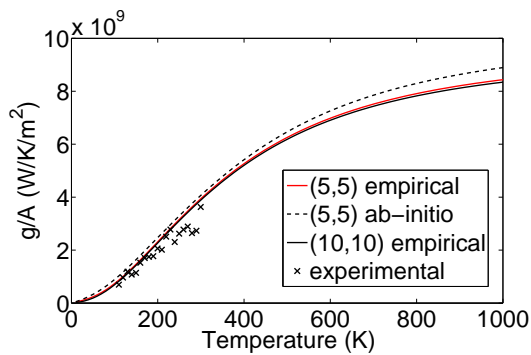}
\caption{\label{fig:ballistic_cond_per_unit_length} (color online) 
  Ballistic conductivity per unit length (expressed in \AA), with the cross sectional
  area~(\ref{eq:cross_sectional_area}) for a (5,5) (red)
  and a (10,10) (black) nanotube. Experimental points taken
  from Ref.~\onlinecite{YSYLM05} are also reported.
}
\end{figure}

Actually, a more intrinsic property is the ballistic thermal  
conductance divided by the cross sectional
surface of the material $g/\mathcal{A}$ (this is the so-called ballistic
conductivity per unit length, compare with Eq.~(\ref{eq:conductivity})).   
The cross sectional area $\mathcal{A}$ is obtained from a ``fattened'' carbon ring: 
\begin{equation}
  \label{eq:cross_sectional_area}
  \mathcal{A} = 2 \pi R d = 3n d,
\end{equation}
where $n$ is the index of the nanotube, $R$ is the
radius of the carbon ring and $d = 3.35$~\AA~
(the graphite interlayer separation)
is some characteristic length 
defining the width of the annular domain enclosing the carbon ring. 
Fig.~\ref{fig:ballistic_cond_per_unit_length}
presents a plot of $g/\mathcal{A}$ as a function
of the temperature, for two CNTs of different indices, as well as a
reference curve computed from {\it ab-initio} results, and
experimental results.\cite{YSYLM05}

Several conclusions can be drawn from this picture:
\begin{enumerate}[(i)]
\item there is a good agreement between the empirical model used and the reference
  results computed with {\it ab-initio} interatomic force
  constants on the whole range of temperatures;
\item the agreement with the experimental results from Ref.~\onlinecite{YSYLM05}
  for temperatures around room temperature and below,
  suggests that the present models contain the essentials ingredients for 
  describing thermal transport. In particular, other effects
  (presently not taken into account) such as 
  anharmonic interactions may be neglected in this
  temperature regime (and for lower temperatures).
  The measurements from Ref.~\onlinecite{YSYLM05} have been
  transformed into conductivities per unit length 
  by assuming that the CNTs used for
  the experiments have a diameter of 1~nm. The
  authors of Ref.~\onlinecite{YSYLM05} indeed precise that the CNTs used
  have a diameter in the range 1-2~nm, occasionally 2-3~nm. 
  If the diameter is indeed 1~nm, then the experimental results 
  are close to the ballistic conductance curve, which means that the
  thermal transport is nearly
  ballistic.\cite{YSYLM05} If the diameters are larger, there is a
  reduction of about 50\% of the conductance 
  for the CNTs of experimental lengths (about
  $L = 2.76~\mu$m) with respect to the ballistic conductance. This
  reduction may be attributed to anharmonic effects. 
  We remark that, even in this case, the effect of isotopic disorder should 
  still be noticeable since the results on the
  conductivities in Section~\ref{sec:thermal_cond_temperature} show that isotope disorder can lead
  to a reduction of about 80\% of the conductance at room temperature
  for CNTs of experimental lengths.
\end{enumerate}
In the sequel, results for temperatures much larger than room
temperature are also presented. We warn however the reader that these
results are given for completeness (having computed the
transmission, the thermal conductance can straightforwardly 
be evaluated at any temperature). 
It is likely that anharmonic effects become more
important as the temperature increases, invalidating the use of a
harmonic description in this temperature range.

\subsubsection*{Parameters used}

The parameters used in this study are given in
Table~\ref{tab:CNT_parameters_5} when the empirical model is used.
In the case of {\it ab-initio} computed IFC, no averaging was performed (since, as shown 
below in Section~\ref{sec:scaling_thermal}, 
the conductance does not change much from one realization of the 
mass disorder to another), but the CNTs considered have the same lengths.
Since the interatomic distance at equilibrium is 
$L_0 \simeq 1.44$~\AA, the nanotubes considered have lengths up 
to~$L=2.49$~$\mu$m, which is a length typical of the currently
available experimental single walled CNTs. The larger system 
({\it i.e.} (10,10) CNT of length $L=2.49$~$\mu$m) consists of $400,000$
atoms.

In all this study, $c=0.5$, and the mass disorder corresponds to
replacing $^{12}$C by $^{13}$C ($\Delta m/m = 1/12$). The qualitative
features presented are robust with respect to a smaller disorder
concentration, but longer tubes should then be studied.
Also, the regularizing parameter 
$\eta/\omega_\text{max}^2 = 10^{-8}$.

For {\it ab-initio} computed IFC, the transmission is computed  
for $N_\omega = 200$ points uniformly spaced
in the range $0 \leq \omega \leq \omega_\text{max} = 1650$~cm$^{-1}$. 
For (5,5) CNTs described by the empirical model, $N_\omega = 1000$ points uniformly spaced
in the range $0 \leq \omega \leq \omega_\text{max} = 1650$~cm$^{-1}$. 
For (10,10) CNTs described by the empirical model, the spectral region $0 \leq \omega \leq
330$~cm$^{-1}$ is discretized using $N_\omega = 200$ points, while 
$N_\omega = 200$ other points are used for the remaining part of the
spectrum. 
Such a decomposition ensures that the low frequency part of
the spectrum, which is very important to have reliable estimates of
the thermal conductivity, is treated accurately, while keeping the
overall computational cost reasonable.
The Boltzmann transmissions are computed in all case with $N_\omega=1000$.

We studied the dependence on the results on the disorder realization, in the case of CNTs 
described by the empirical model.
For each tube length $L$, $N_\text{disorder}$ realizations of the
isotopic disorder are considered, and the transmission are averaged
over the different realizations. We considered a fixed
computational cost for each tube length, so that $L N_\text{disorder}$
is constant. 

\begin{table}
\caption{ \label{tab:CNT_parameters_5} Number of averages over the mass
  disorder realizations as a function of nanotube length for CNTs described by the 
  empirical model.}
\begin{ruledtabular}
\begin{tabular}{@{}ccccccc@{}} 
Length (nm) &  2490 & 823 & 249 & 82.3 & 
24.9 & 8.23\\
\hline
$N_\text{layers}$ & $10^4$ & 3300 & 1000 & 330
& 100 & 33 \\
$N_\text{disorder}$ & 1 & 3 & 10 & 30 & 100 & 300 \\
\end{tabular}
\end{ruledtabular}
\end{table}


\subsection{Average transmissions}
\label{sec:average_transmissions}

\begin{figure}[htbp]
\center
\includegraphics[width=7.3cm]{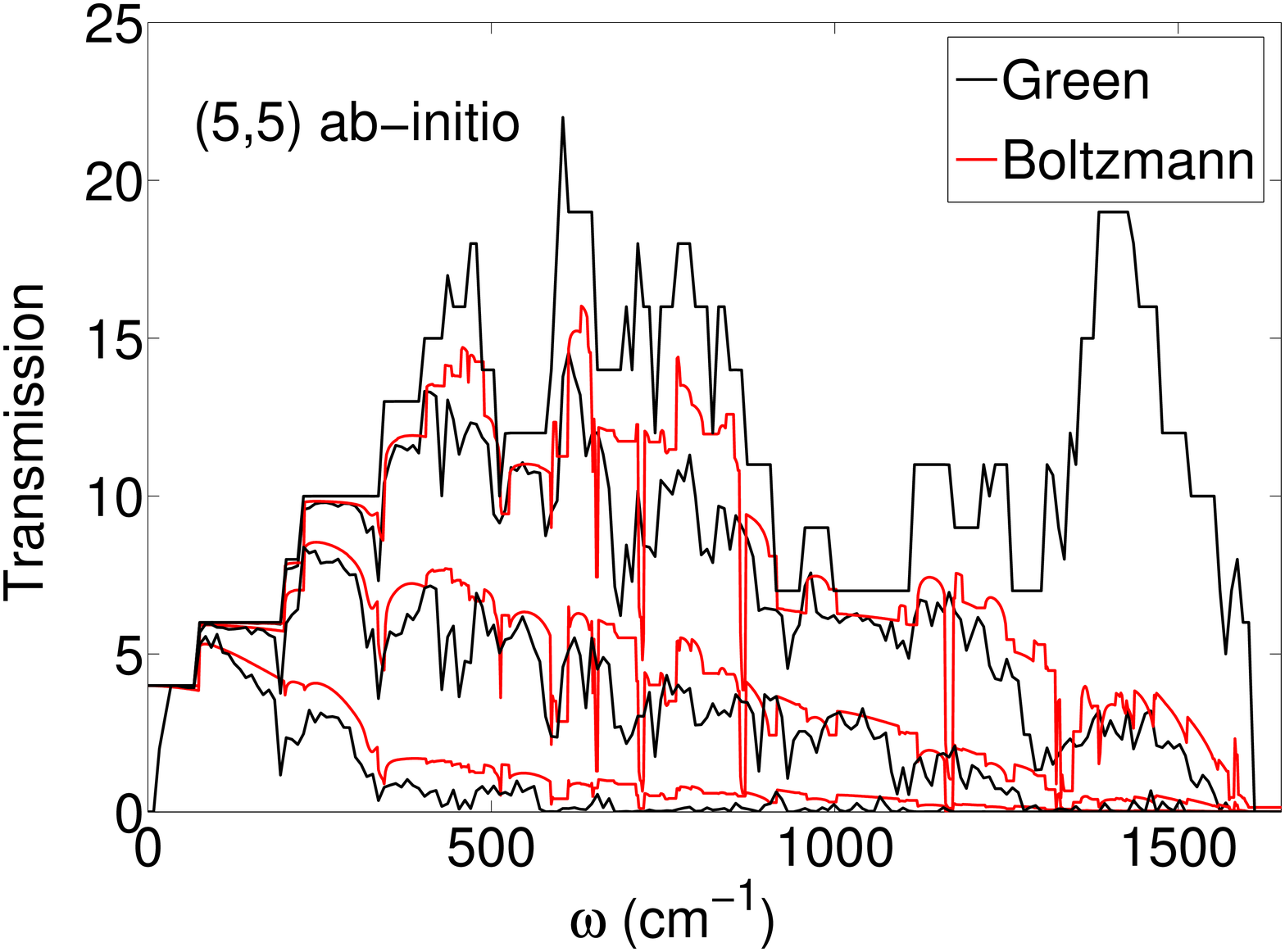} 
\includegraphics[width=7.3cm]{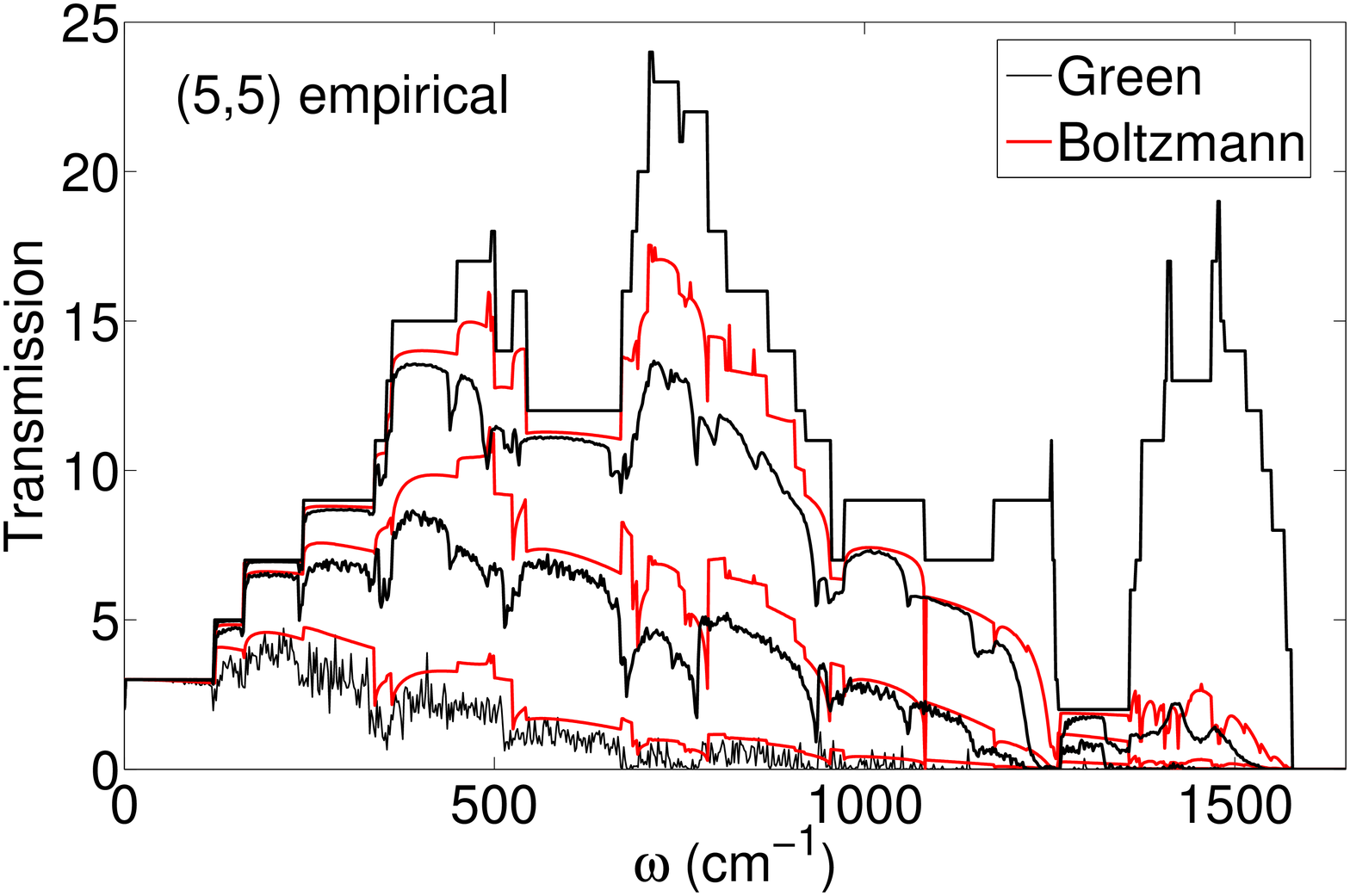} 
\includegraphics[width=7.3cm]{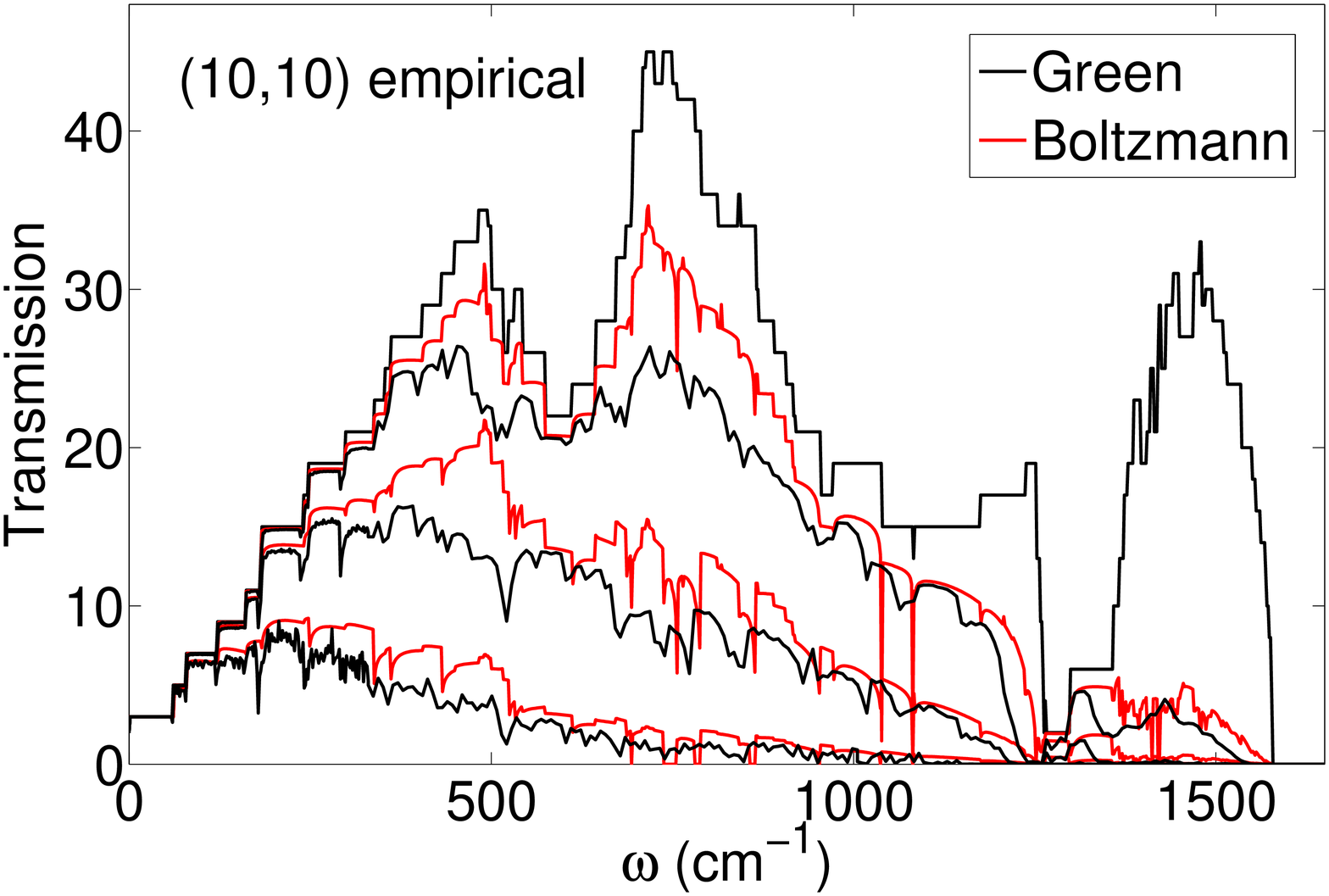} 
\caption{\label{fig:trans_CNT} (color online) 
  Transmissions obtained with the Green's function formalism (black
  curves) and the Boltzmann approach (red curves). In each of the three figures, 
  the line corresponding to the largest transmission is the reference
  ballistic transmission. The other lines correspond to tubes of
  different lengths, namely $L=25$~nm, $L=249$~nm and $L=2.49$~$\mu$m,
  the highest transmissions corresponding to the shortest tubes.
}
\end{figure}

The average transmission functions are presented in 
Fig.~\ref{fig:trans_CNT}. 
The different transmission pictures obtained show that the higher
frequency modes are damped out first, while the acoustic modes are
less perturbed. In general, the transmission decreases with the
pulsation. 
Longer system sizes 
are necessary to modify substantially the transmission of acoustic modes. 
As the carbon nanotube index (equivalently, its diameter)
increases, the spikes in the transmission become less
marked, and the behavior of the transmission as a function of the
pulsation is almost monotonic.

Although not shown here, the maximal variance over the transmission
functions computed for different realizations of the mass disorder
is constant for different tube lengths, as is the case for
one-dimensional chains. However, the {\it relative} variation of the
transmission (defined
as the square root of the variance divided by the average transmission)
decreases with the CNT index. This is due to the fact that
transmission fluctuations remains of order 1, while the ballistic
transmission (the number of conducting channels) 
increases proportionally to the system index.

The Boltzmann transmission is in fair agreement with the Green's
function transmission, especially at low frequencies. At higher
frequencies, the Boltzmann transmission is usually higher than the
Green's function one. This is in
analogy with the results for the dimensional chains.
The largest discrepancies arise however for pulsations in the 
range $\omega = 400 - 500$~cm$^{-1}$, corresponding to the
out-of-plane phonon modes. 
The agreement seems however better when {\it ab-initio} IFCs are used. 
This may be due to the fact that in-plane and out-of-plane modes interact 
in this case (indeed, the
Boltzmann treatment is expected to be more accurate when the number of
interacting modes is larger).


\subsection{Determination of the transmission regime at some chosen
  frequencies}

Anderson localization\cite{Anderson58} is often observed in one-dimensional disordered
systems. In this case, the transmission function decreases
exponentially with the system size (in average). This is the case for 
isotopically disordered harmonic one-dimensional chain (see
Section~\ref{sec:theoretical_heat_harmonic_isotopic}), for which
however the localization length goes to infinity when the pulsation
goes to 0. When many conducting channels are present in a disordered
harmonic system, it is unclear
whether the transmission is still localized, or if it becomes closer to a
diffusive behavior.

We discuss here whether the transmission regime is diffusive or
localized for a given pulsation~$\omega$. There are several ways to
determine the nature of the transmission regime, 
see for instance the recent work\cite{ARTBL07} where diffusive 
and localized lengths for different system sizes are computed.
We compute here the diffusive and localized lengths
by resorting to some functional \textit{ansatz} of the transmission
function which allows to interpolate between the two regimes. The
transmission considered is the average transmission since this is the
quantity of interest.
The (average) transmission function computed by the Green's function
and the Boltzmann approaches
may be approximated by the following \textit{ansatz}:
\begin{equation}
  \label{eq:ansatz_asymptotic_transmission}
  T^\text{ansatz}_L(\omega) = T_\textrm{ballistic}(\omega) \, 
  \left ( 1 + \frac{L}{l_\textrm{diff}(\omega)} 
  \right)^{-1} \text{e}^{-L/l_\textrm{loc}(\omega)},
\end{equation}
where $l_\textrm{diff}(\omega)$ is some characteristic length
associated with diffusive transport at the pulsation $\omega$
(inducing a $1/L$ decay) while 
$l_\textrm{loc}(\omega)$ measures the propensity of the system to 
localize its states (inducing an exponential decay of the transmission
as in the Anderson model). 

To determine the lengths $l_\textrm{diff}(\omega),
l_\textrm{loc}(\omega)$, we compute $N_\text{realizations}$ 
values of the transmission at a fixed value of $\omega$ for several 
lengths $L=L_1,\dots,L_{N_\text{lengths}}$, and perform a
least-square fit of the average transmission based on the
\textit{ansatz}~(\ref{eq:ansatz_asymptotic_transmission}). 
Many realizations should be considered in order to have
reliable results, and so, we restricted ourselves to the simple empirical model; 
here $N_\text{realizations}=300$. 
Besides, it is actually not always sufficient to perform a 
least-square fit directly on the transmissions. Indeed, when
the transmission regime is (close to) localized, the direct estimates
give poor results since the transmission is asymptotically very
small, and the corresponding points have very small weights in the 
direct least-square fitting procedure. 
Therefore, we resort to 
a nonlinear least-square fit, where $f(T_L(\omega))$ and 
$f(T_L^\text{ansatz}(\omega))$ are compared.
The function $f$ puts more emphasis on the asymptotic regime, 
{\it i.e.} on the smaller values of the transmission. The choice 
$f=-\log$ is convenient, but other
choices such as $f(x)=1/x$ yielded similar results.

\begin{figure}[htbp]
\center
\includegraphics[width=7.3cm]{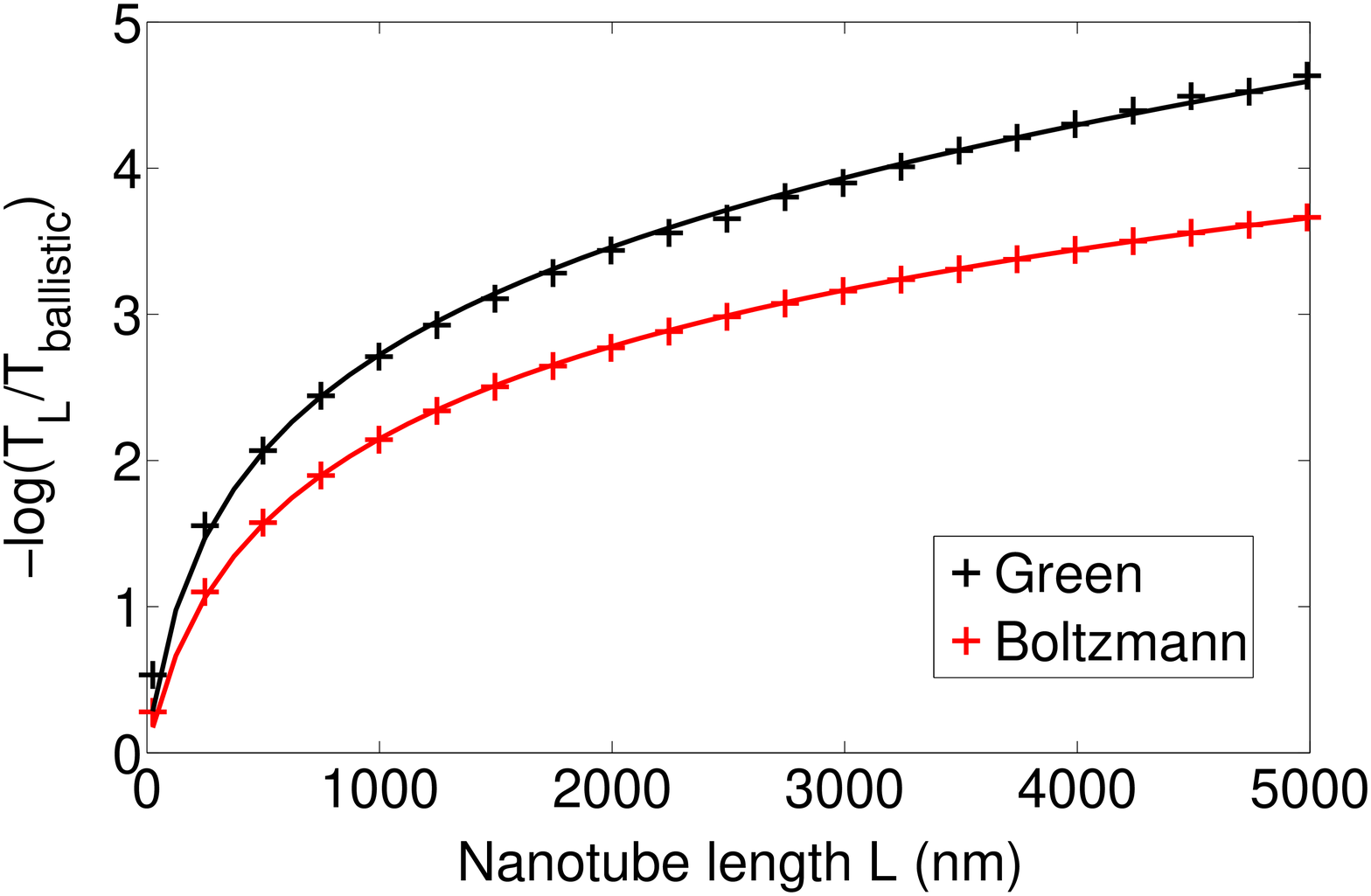}  
\caption{\label{fig:fit_ansatz} (color online) 
  Fit of $-\ln( T_L/T_\text{ballistic} )$ as a function of $L$ at
  $\omega = 716$~cm$^{-1}$ for a (10,10) CNT described by the empirical model. 
  The transmission computed with the Green's function and Boltzmann methods
  can be fitted with a good precision using the 
  ansatz~(\ref{eq:ansatz_asymptotic_transmission}) (solid black and
  red lines).
}
\end{figure}

Fig.~\ref{fig:fit_ansatz} presents a typical fit in the
diffusive regime at the frequency $\omega = 716$~cm$^{-1}$ for a (10,10)
CNT, and compares the normalized transmissions
$T_L(\omega)/T_\text{ballistic}(\omega)$ computed using both Boltzmann
and Green's function methods. As can be seen, the functional form of
the {\it ansatz}~(\ref{eq:ansatz_asymptotic_transmission})
is indeed appropriate.

The characteristic lengths as for different pulsations
are presented in Table~\ref{tab:diffusive_lengths} for
(5,5) CNTs, and in Table~\ref{tab:diffusive_lengths_10} for
(10,10) CNTs (for some pulsations, the transmission decreases
too fast with increasing system size, and no fitting could be performed).
Using these results, it is possible to determine whether the
transmission regime is diffusive or localized.
Notice however that 
the diffusive and localized lengths as defined 
by~(\ref{eq:ansatz_asymptotic_transmission}) should not be compared
directly when they are of the same order of magnitude. 
Rather, given some attenuation factor $0 \leq \gamma \leq 1$
(typically, $\gamma=0.1$), one should compare the lengths of the sample
required to reduce the ballistic transmission by a factor $\gamma$. For the
diffusive term, this is $L \sim (1/\gamma - 1) l_\textrm{diff}(\omega)$, while
for the localized regime $L \sim - l_\textrm{loc}(\omega) \ln \gamma $.

The results of Tables~\ref{tab:diffusive_lengths} and~\ref{tab:diffusive_lengths_10}
show that, for the exact (Green's function) transmission,
the localization length decreases with increasing frequencies, so that localization effects
become more and more important as the pulsation increases. This
explains the fast decrease of the high frequency modes. The behavior
of the diffusive lengths is not as clear cut as the behavior of the
localization lengths: the diffusive lengths
decrease at smaller pulsations, and then settle down to values around
$100-300$~nm, which are therefore completely relevant for CNTs of
experimental sizes. It seems that some minimal diffusive length is
obtained when the number of channel is maximal, which is consistent
with the picture that diffusive transport arises from random scattering
between many channels.

For the Boltzmann transmission, only the diffusive
part is relevant since the localization lengths computed are always
much larger than the diffusive lengths. There is also a trend towards
shorter diffusive lengths as the pulsation increases. 
This may be explained by the fact that the scattering for 
higher frequencies is much more efficient than for acoustic modes,
which, in turn, comes from the $\omega^2$ dependence in the expression
of the scattering matrix \eqref{eq:facteur_isotopic_disorder}.

Finally, it is
observed that both diffusive and localization lengths increase with
increasing diameter of the system, especially the localization
length. This can related again to an increase of the number of
available modes with increasing diameters, and thus, a reduced
scattering. 

\begin{table}
\caption{ \label{tab:diffusive_lengths} Localization and diffusive
  lengths (expressed in nm) for a (5,5) CNT described by the empirical model. 
}
\begin{ruledtabular}
\begin{tabular}{cccccc} 
& & \multicolumn{2}{c}{Green} & \multicolumn{2}{c}{Boltzmann}\\
$\omega$ (cm$^{-1}$) & $T_\text{ballistic}$ & $l_\textrm{diff}$
& $l_\textrm{loc}$ & $l_\textrm{diff}$ & $l_\textrm{loc}$\\
\hline
90 & 3 & 77100 & $7 \times 10^7$ 
& $1 \times 10^5$ & $1 \times 10^7$\\
179 & 7 & 3399 & $2 \times 10^{7}$ 
& 5273 & 456000\\
269 & 9 & 1422 & $1 \times 10^{11}$ 
& 2990 & $9 \times 10^{10}$ \\
358 & 13 & 304 & 52800 & 897& $2 \times 10^{10}$\\
448 & 15 & 387 & $2 \times 10^{11}$ & 779 & 212000 \\
537 & 16 & 239 & 13700 & 433 & 33800 \\
627 & 12 & 320 & 10900 & 455 & 28100 \\
716 & 23 & 72.9 & 4897 & 139 & 24600 \\
806 & 18 & 119 & 6552 & 215 & 27100 \\ 
896 & 14 & 101 & 4007 & 180 & 23900 \\
985 & 9 & 149 & 2502 & 156 & 23800 \\
1075 & 9 & 117 & 1716 & 85 & 19600 \\
1164 & 7 & 341 & 289 & 78.2 & 18300 \\
1254 & 7 & $2 \times 10^6$ & 207 & 2.62 & 1993 \\
1343 & 2 & $9 \times 10^9$ & 39.8 & 288 & 25200\\
1433 & 13 & $2 \times 10^{10}$ & 136 & 6.35 & 4392 \\
\end{tabular}
\end{ruledtabular}
\end{table}

\begin{table}
\caption{ \label{tab:diffusive_lengths_10} Localization and diffusive
  lengths (expressed in nm) for a (10,10) CNT described by the empirical model.
}
\begin{ruledtabular}
\begin{tabular}{cccccc} 
& & \multicolumn{2}{c}{Green} & \multicolumn{2}{c}{Boltzmann}\\
$\omega$ (cm$^{-1}$) & $T_\text{ballistic}$ & $l_\textrm{diff}$
& $l_\textrm{loc}$ & $l_\textrm{diff}$ & $l_\textrm{loc}$ \\
\hline
179 & 11 & 1927 & $4 \times 10^{10}$& 6727 &
$1 \times 10^6$ \\
358 & 25 & 494 & $1 \times 10^{12}$ & 1165 & 
241000 \\
537 & 30 & 235 & 212000 & 395 & 31900\\
716 & 45 & 76.5 & 12500 & 166 & 
26000 \\
896 & 26 & 125 & 9269 & 194 & 24200 \\
1075 & 15 & 101 & 2851 & 118 & 21300 \\
1254 & 13 & -- & -- & 3.12 & 1824 \\
1433 & 29 & 114 & 88.0 & 5.43 & 3189 \\
\end{tabular}
\end{ruledtabular}
\end{table}


\subsection{Asymptotic scaling of the thermal conductance}
\label{sec:scaling_thermal}

\begin{figure}[htbp]
\center
\includegraphics[angle=0,width=7.3cm]{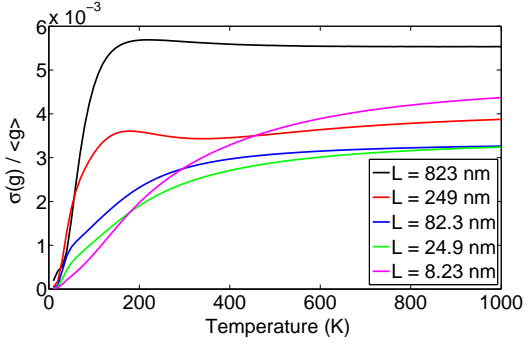}
\caption{\label{fig:Variance_CNT_5} (color online)
  Standard deviation of the conductance $\sigma(g)$ of a (5,5) CNT
  (described by the empirical model) divided by average conductance $\langle g \rangle$, 
  as a function of temperature, for different system sizes.
} 
\end{figure}

Before discussing the properties of the thermal conductance, we first check whether
the realizations of the mass disorder have a noticeable influence on
the value of the normalized thermal conductance $\overline{g}$ given
by~(\ref{eq:normalized_thermal_conductance}). This is important from an
experimental viewpoint since only few (if not only one) realizations of the
mass disorder can be considered. 
Fig.~\ref{fig:Variance_CNT_5} presents the normalized standard
deviation of the conductance, for different system sizes, as a function of the
temperature. The variations of the
conductance are lower than 1\% from one realization to the other. This
is due to the fact that the conductance is the integral of the
transmission function (with the proper weighting function), and
therefore the variations of the transmission function for
only one realization of the mass disorder are averaged out.

\begin{figure}[htbp]
\center
\includegraphics[width=7.3cm]{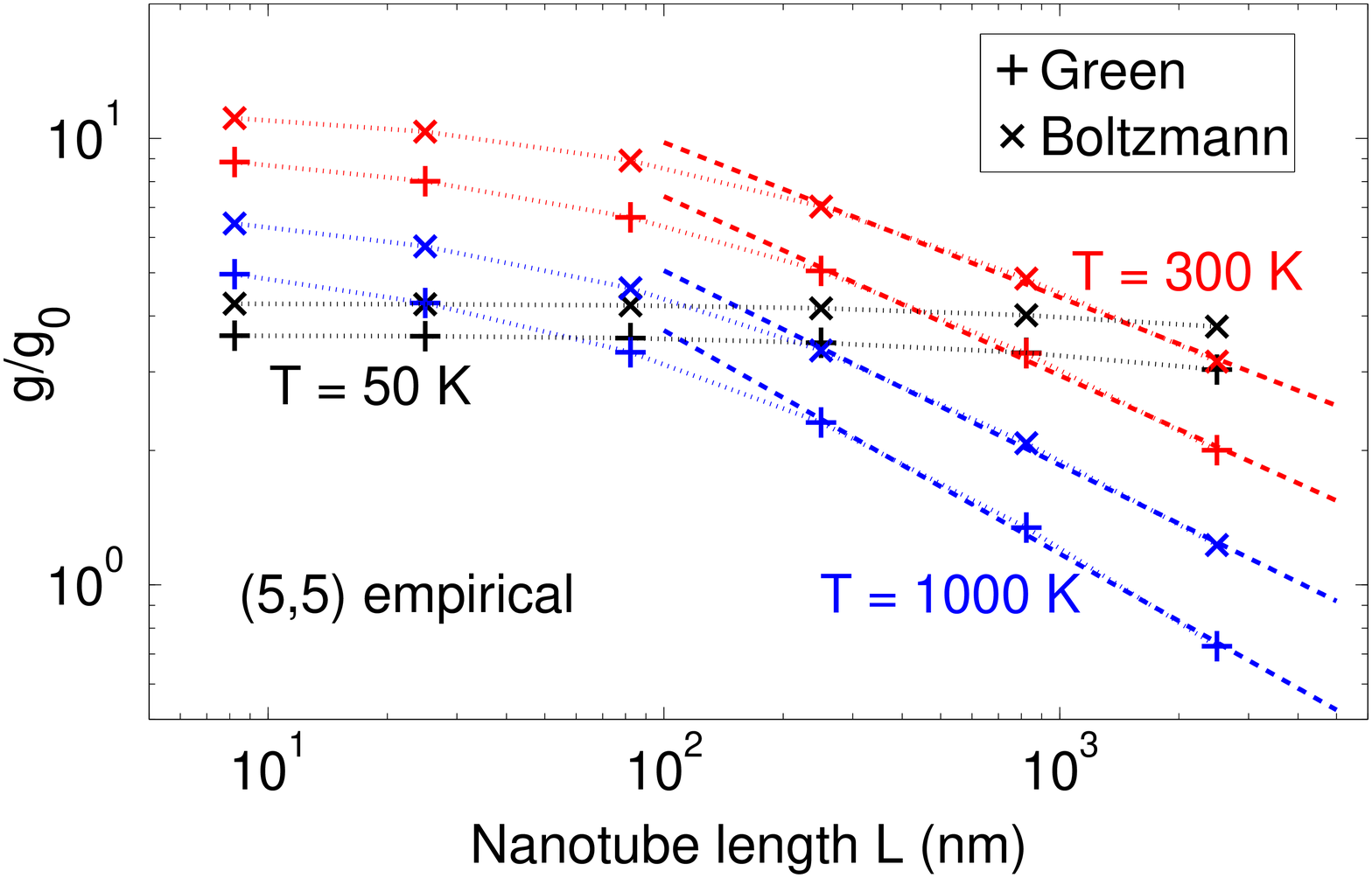}
\includegraphics[width=7.3cm]{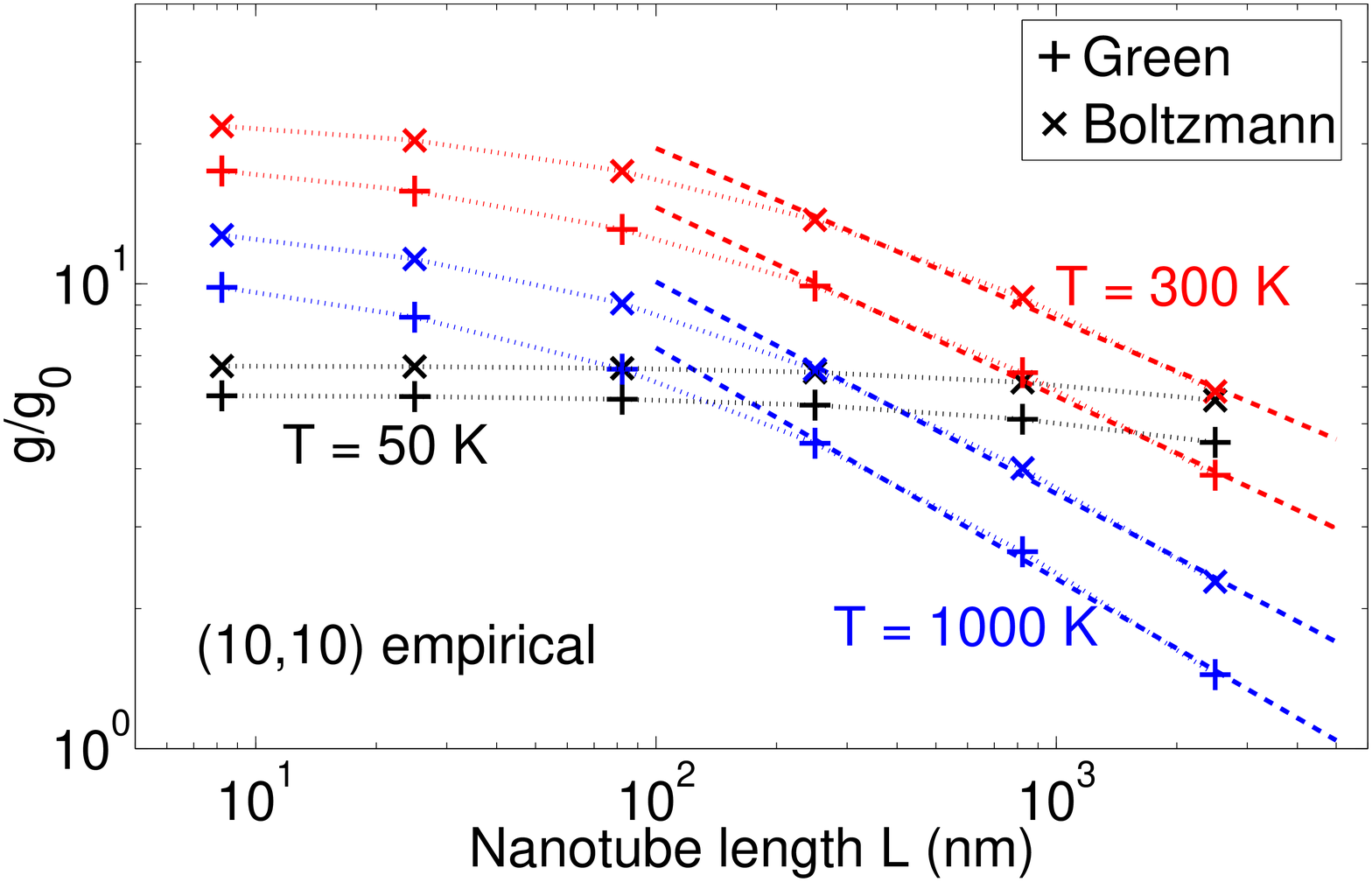}
\includegraphics[width=7.3cm]{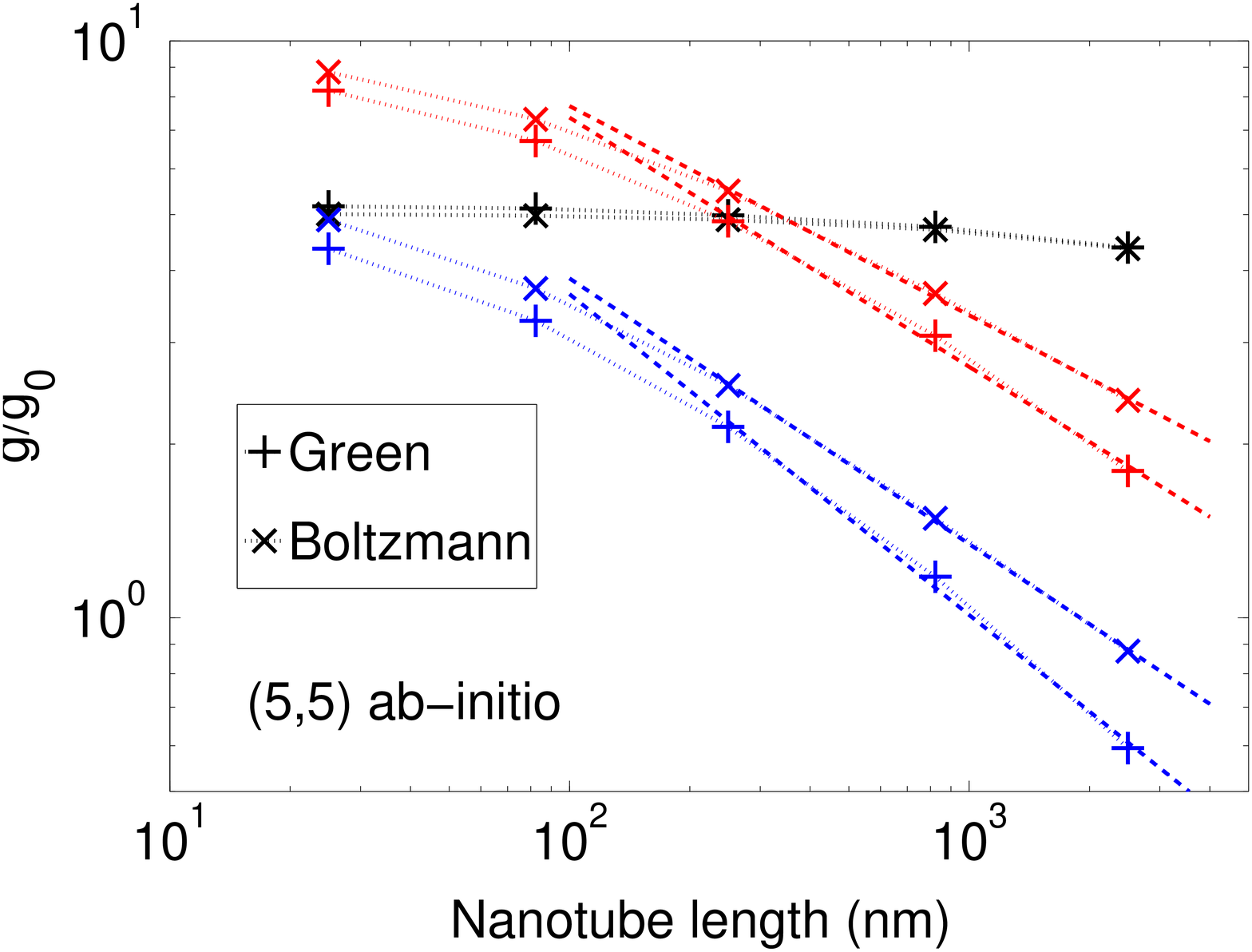}
\caption{\label{fig:conductance_scaling} (color online)
  Variation of the normalized conductance 
  for temperatures $T=50$~K (black curves), 
  $T=300$~K (red curves), $T=1000$~K (blue curves), for the
  conductivities computed with the Green's function approach and the
  conductivities computed with the Boltzmann approach. For a given
  temperature, the Boltzmann curve is always above the Green's
  function curve. 
  Top: (5,5) CNT, estimated $\alpha=0.40$ at $T=300$~K and 
  $\alpha=0.50$ at $T=1000$~K.
  Middle: (10,10) CNT, estimated $\alpha=0.41$ at $T=300$~K and 
  $\alpha=0.50$ at $T=1000$~K.
  Bottom: (5,5) CNT with {\it ab-initio} IFCs, estimated $\alpha=0.43$ at $T=300$~K and 
  $\alpha=0.55$ at $T=1000$~K.
}
\end{figure}

Fig.~\ref{fig:conductance_scaling} presents the 
averaged normalized thermal conductance as a function of the system length (in log-log
scale) for (5,5) and (10,10) CNTs. 
Three temperatures are considered in both cases:
$T=50$~K, $T=300$~K, $T=1000$~K. 
The scalings obtained for $T=50$~K show that the transmission regime
is quasi-ballistic since the conductance does not change much with the
system size. Indeed, at low temperatures, only the low frequencies
modes, which are almost unaffected by the disorder, matter.
For higher temperatures, more and more higher frequency modes are 
introduced, so that
disorder has a noticeable effect, and the
conductance decreases as a function of the system size.
When this makes sense, the slope
$\alpha$ of the different curves have been estimated using some
least-square fitting, in order to characterize a power law decay 
$\overline{g} \sim L^{-\alpha}$. The exponents found are in the
range $\alpha = 0.4 - 0.5$ at $T = 300$~K. The associated thermal conductivities
therefore have a power-law divergence 
$g L/\mathcal{A} \sim L^\beta$ with $\beta = 1-\alpha = 0.5 - 0.6$ for
the range of lengths considered in this study.
The general trends in the curves are quite comparable
for (5,5) and (10,10) CNTs, independently of the model used.

Notice that the asymptotic regime is not yet attained for nanotubes
of lengths up to $2.5~\mu$m (which are typical experimental lengths). 
In this regime the exponent $\alpha$ is not universal
since it depends on the tube length and on the amount of mass
disorder.  
As in the case of
one dimensional chains, the truly
asymptotic regime corresponds to $\overline{g} \sim L^{-1/2}$ (or
$\kappa \sim L^{1/2}$),
with associated transmission profiles
where only acoustic modes have a non-zero transmission. To
this end, much larger system sizes (or a larger mass
disorder) should be considered.
In any cases, it is observed that there is no well-defined
conductivity.


\subsection{Thermal conductance as a function of the temperature}
\label{sec:thermal_cond_temperature}

\begin{figure}[htbp]
\center
\includegraphics[width=7.3cm]{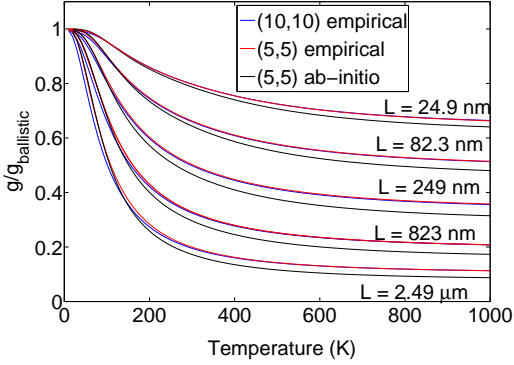}
\caption{\label{fig:conductance_temperature_ratio} (color online)
  Ratio of the thermal conductance to the ballistic
  conductance, for (10,10) empirical CNT (blue curves), 
  (5,5) empirical CNT (red curves) and (5,5) ab-initio CNT (black curves). 
  From top to down: increasing tube lengths.
}
\end{figure}

The temperature dependence of the 
normalized conductance is
presented in Fig.~\ref{fig:conductance_temperature_ratio}. 
Those curves are obtained for the
transmission computed with the Green's function approach, and the
curves for the transmission computed using a Boltzmann treatment have
a very similar behavior with respect to temperature.
As expected, the ratio of the thermal conductance to the
ballistic one is always smaller than~1, and converges to~1 in the low
temperature limit. 

Several conclusions can be drawn from this picture:
(i) the relative reduction of the conductance due to the
isotope disorder does not depend on the CNT diameter;
(ii) the decrease predicted by Boltzmann equation is in excellent
agreement with the decrease found with the Green's function method;
(iii) isotopic disorder can be very efficient in reducing the 
thermal conductivity,
especially for tubes of experimental lengths, even at moderately high
temperatures. For instance, for (5,5) and (10,10) tubes of experimental
lengths ($L=2.49$~$\mu$m), 
the thermal conductance is decreased by 80\% at room temperature.
Notice that the decrease in the thermal conductivity increases with
the temperature. This is consistent with the results of the previous
sections since, as the temperature is increased, more and more 
higher frequency modes are introduced, and the transmission function
is almost a decreasing function of the pulsation.

\begin{figure}[htbp]
\center
\includegraphics[width=7.3cm]{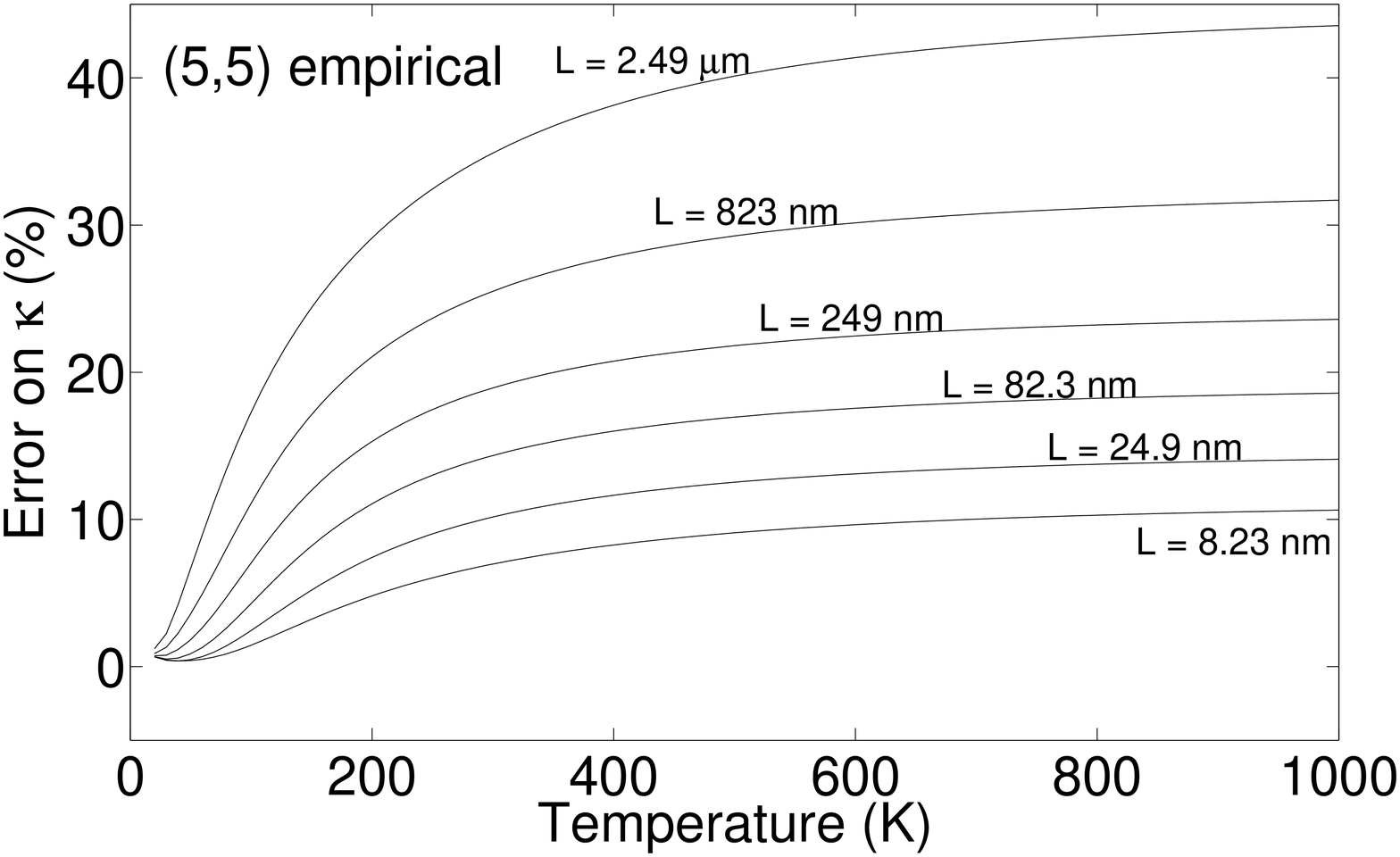} 
\includegraphics[width=7.3cm]{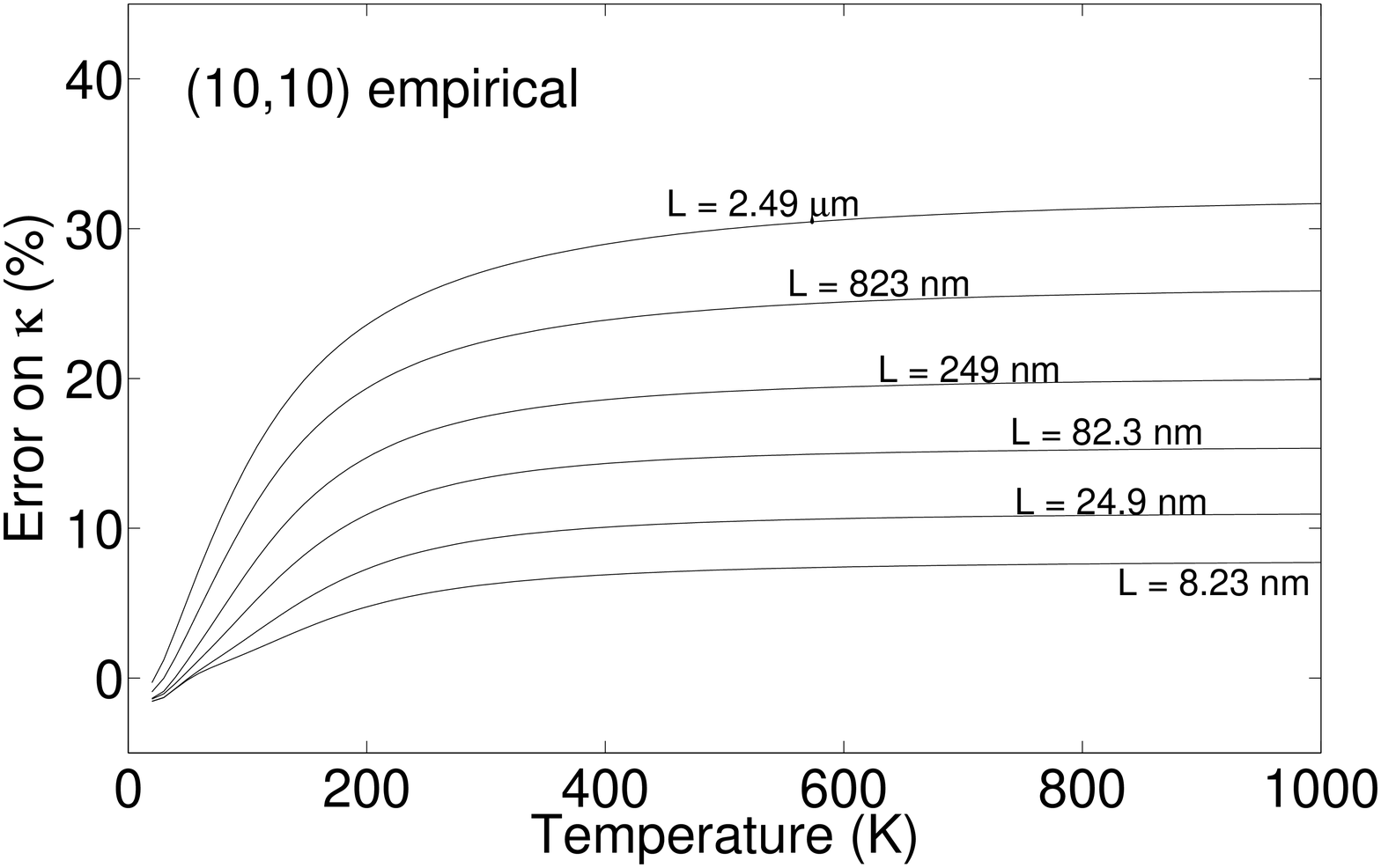} 
\includegraphics[width=7.3cm]{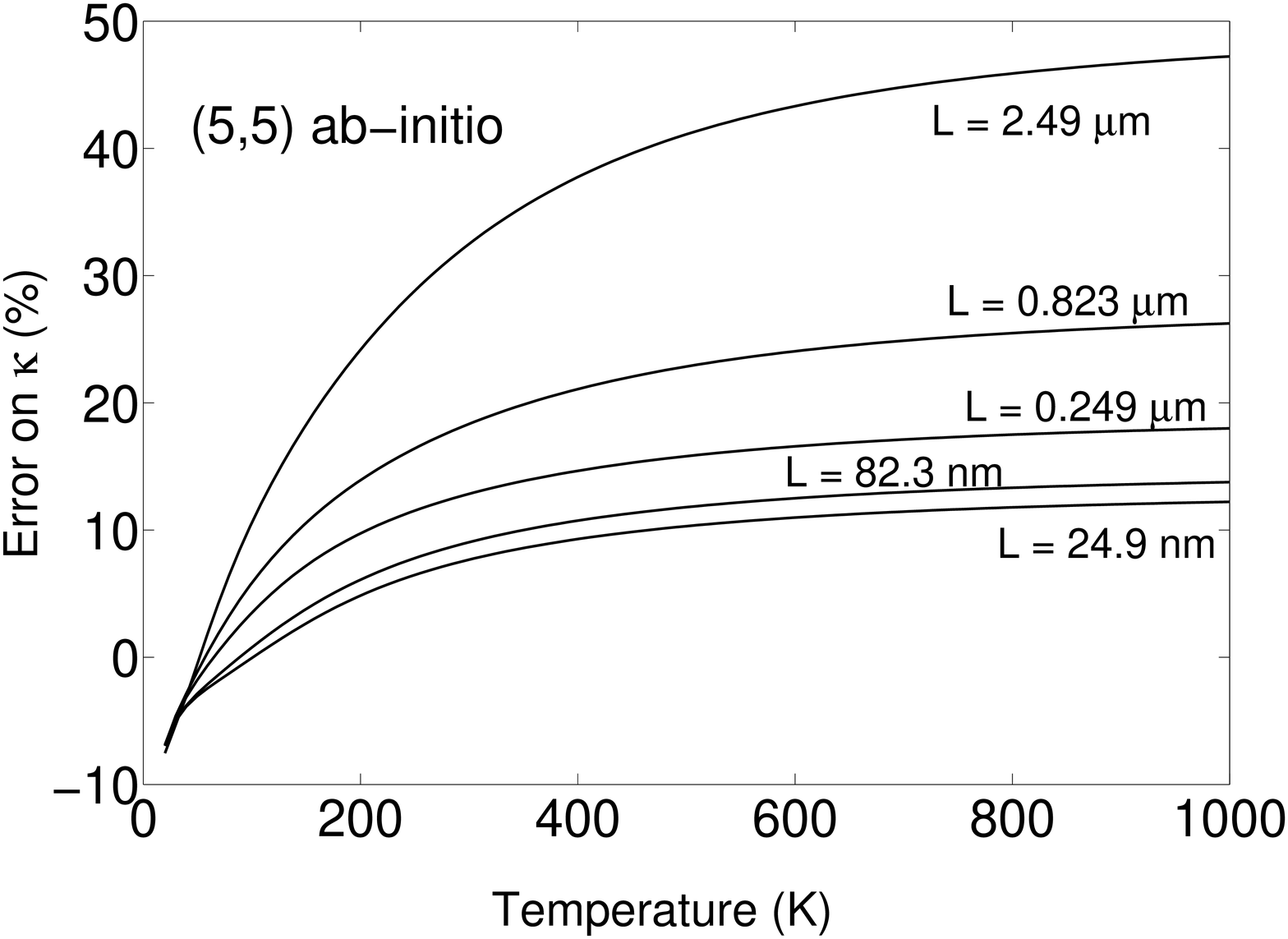} 
\caption{\label{fig:comp_boltz_conductivity}
  Error on the the conductivity computed with the profile
  obtained from the Boltzmann approach. The reference value is
  computed from the profiles obtained with the Green's function
  approach (decreasing lengths from top to down). 
}
\end{figure}

The thermal conductivity (or conductance) 
predicted by the Boltzmann approach is compared
to the conductivity (or conductance) obtained by the Green's function approach. The
results are presented in Fig.~\ref{fig:comp_boltz_conductivity},
where the error 
\[
e(T) = \frac{\kappa_\text{Boltzmann} - \kappa_\text{Green}}{\kappa_\text{Green}},
\]
is plotted as a function of the temperature $T$. As can be seen, 
the Boltzmann approach gives correct orders of magnitude in
the temperature range and for the CNT index considered here.
It however overestimates the conductivity, and 
the error increases with the length of the system and the
temperature. This can be explained by the fact that the Boltzmann
approach is not precise enough to capture the real decay of the
transmission function. In particular, the decrease of the transmission
is not fast enough, which is consistent with the results of
Fig.~\ref{fig:trans_CNT}. 

However, as already pointed out in 
Section~\ref{sec:average_transmissions}, 
it is expected that the Boltzmann treatment becomes more
accurate as the number of interacting modes increases. It is therefore
likely that the agreement between the conductance computed using 
Green's function techniques and conductances computed from the
Boltzmann transmissions will be better for 
tubes with larger indices or multi-walled tubes (or for models
treating the out-of-plane modes more precisely).
This is indeed true, as can be seen from the results presented in 
Fig.~\ref{fig:comp_boltz_conductivity}, where 
the error on the conductivity is smaller for
CNTs of larger diameters (To give some orders of magnitude, the
diameters of (5,5) and (10,10) CNTs are respectively 0.69~nm and
1.38~nm). 

%
%

\section{Conclusion}
\label{sec:conclusion}

We studied the thermal transport of isotopically disordered 
harmonic CNTs using Green's function and Boltzmann treatments.
The Green's function method is implemented using an algorithm, whose
computational cost scales linearly with the system size.  
It is not restricted to isotopic disorder, and can be applied to other
kind of defects which scatter phonons elastically.\cite{MSBS08} In view of
the theoretical results for 1D systems, we also 
expect a divergence of the thermal conductivity with the
system size in those cases.

Our numerical results show that
\begin{enumerate}[(i)]
\item these systems share some common
  features with the thermal transport in isotopically disordered
  harmonic one dimensional atom
  chains, in particular a power law divergence of the thermal
  conductivity $\kappa \sim L^\alpha$. 
  Therefore, Fourier's law is not valid.
  For tubes of experimental length, the exponent of the power law
  divergence $\alpha \simeq 0.6$ at room temperature. For longer tubes,
  the exponent may well be different since $\alpha = 1/2$ in the asymptotic regime
  where only the acoustic branches have a non-zero transmission.
  Experimental results on 
  the length dependence of disordered CNT conductivities, 
  measuring divergences $\alpha \simeq 0.4-0.6$ for boron nitride
  tubes (with a large fraction of mass disorder),  
  have been published in Ref.~\onlinecite{COGMZ08}
\item the thermal conductivity decreases monotonically with the temperature, 
  and is independent of the CNT diameter. 
  We showed that there is a dramatic reduction of the thermal
  conductance for systems of experimental sizes (roughly 80\% at room
  temperature), when a large fraction
  of isotopic disorder is introduced. 
  This is in accordance with
  experimental measurements of the effect of isotope disordered in boron
  nitride nanotubes, which demonstrated dramatic changes in the 
  conductances (a conductance enhancement by 50\% for purified 
  materials);\cite{CFAOIG06}
\item a Boltzmann description of the
  thermal transport gives correct results for the thermal
  conductance even in the presence of Anderson localization.
  This is particularly interesting since the computation of the
  transmission using Boltzmann's equation is much less computationally
  expensive, so that larger systems (such as multi-walled CNTs or boron
  nitride nanotubes, or single-walled CNTs with a large diameter) 
  may be studied with this method.
  This shows also that Anderson localization, 
  that is not accounted for in the Boltzmann approach, 
  does not have a clear signature in thermal
  transport for CNTs of experimental sizes; 
\item the results discussed in the previous points 
  are robust with respect to the model used to describe the phonon
  dispersion in the nanotube. In particular, the behaviors predicted by a simple
  empirical model which neglects the tube curvature are very similar to
  those obtained with {\it ab-initio} force constants, computed 
  with density functional theory, and considering the tube curvature.
\end{enumerate}

\begin{acknowledgments}
We thank Nicola Bonini for the {\it ab-initio} computed 
interatomic force constants used in this study. Helpful
discussions with Nicola Bonini and Nicola Marzari are also 
acknowledged. 
Part of this work was done while G.~Stoltz was participating to the
program ``Computational Mathematics'' at the Haussdorff Institute for
Mathematics in Bonn, Germany.
\end{acknowledgments}

%
%

\appendix


\section{Computation of the transmission within the Green's function formalism}
\label{sec:Green_num}

In order to simplify the notations, the number of layers in the
disordered region $N_\text{layers}$ will be denoted by 
$N$ in this section.
Besides, the parameter $\eta$ is always kept positive 
in numerical computations. We indicate explicitly the dependence on
this parameter in this section.

The $(i,j)$ block component of some matrix $\mathcal{M}$ is denoted by
$[\mathcal{M}]_{i,j}$. It is easily seen that the only non zero block
components of the self-energies are respectively 
$[\Sigma_L^\eta(\omega)]_{1,1}$ and $[\Sigma_R^\eta(\omega)]_{N,N}$.
These coefficients are computed 
using standard decimation techniques.\cite{GdSK81,GTFL83,LLR84,LLR85}

The transmission $\cT(\omega)$ can be evaluated using only 
the block $[G^\eta_\text{sys}(\omega)]_{1,N}$ of the full matrix 
$G^\eta_\text{sys}(\omega)$ since
\begin{widetext}
\[
\cT(\omega) = 4 \, \tr \left [ \textrm{Im}([\Sigma_L^\eta(\omega)]_{1,1}) 
\, [G_\textrm{sys}^\eta(\omega)]_{1,N} \, 
\textrm{Im}([\Sigma_R^\eta(\omega)]_{N,N}) 
\left ( [G_\textrm{sys}^\eta(\omega)]_{1,N} \right )^\dagger \right ],
\]
\end{widetext}
where the trace is taken over
a space of dimension $3N_\text{at} \times 3N_\text{at}$. 

The submatrix $[G^\eta_\text{sys}(\omega)]_{1,N}$ can be computed by
solving
\[
G^\eta_\text{sys}(\omega)^{-1}
\left ( \begin{array}{l} 
\left[G_\textrm{sys}^\eta(\omega)\right]_{1,N} \\
\qquad \vdots \\
\left[G_\textrm{sys}^\eta(\omega)\right]_{N-1,N} \\
\left[G_\textrm{sys}^\eta(\omega)\right]_{N,N} \\
\end{array} \right ) = \left ( \begin{array}{c} 
0 \\
\vdots \\
0 \\
\textrm{I} \\
\end{array} \right ),
\]
with 
\[
G^\eta_\text{sys}(\omega)^{-1} = \omega^2+\ri\eta-A_{\rm
  sys}-\Sigma^\eta_L(\omega)-\Sigma^\eta_R(\omega).
\]
We use a
method to compute linear scalingly the last column of the inverse of a
block tridiagonal matrix, based on gaussian elimination. Each
$[G_\textrm{sys}^\eta(\omega)]_{i,N}$ is expressed in terms of 
$[G_\textrm{sys}^\eta(\omega)]_{i+1,N}$ as 
$[G_\textrm{sys}^\eta(\omega)]_{i,N} = F_i [G_\textrm{sys}^\eta(\omega)]_{i+1,N}$.
Denoting by $B = \omega^2+\ri\eta-
A_\text{sys}-\Sigma^\eta_L(\omega)-\Sigma^\eta_R(\omega)$ and 
$f_i =[G_\textrm{sys}^\eta(\omega)]_{i,N} $,
it holds, for $2 \leq i \leq N-1$,
\[
B_{i,i-1} f_{i-1} + B_{i,i} f_i + B_{i,i+1} f_{i+1} = 0, 
\qquad
f_i = F_i f_{i+1}.
\]
It is then easily shown that 
\[
F_1 = -B_{1,1}^{-1} B_{1,2},
\]
\[
F_{i} = -(B_{i,i}+B_{i,i-1}F_{i-1})^{-1} B_{i,i+1} \ \ (2 \leq i \leq N-1)
\]
while $f_N = (B_{N,N} + B_{N,N-1}F_{N-1})^{-1}$, so that 
$[G^\eta_\text{sys}(\omega)]_{1,N} = f_1 = F_1 F_2 \dots F_{N-1} f_N$
is recovered in $\text{O}(Nd^3)$
operations. The corresponding algorithm therefore has a cost
$\text{O}(N_\text{layers} N_\text{at}^3)$ (see Algorithm~\ref{algo:diag_G1N}).

\begin{figure}[htbp]
\bookbox{
\begin{center}{\sc Iterative computation of the $(1,N)$ block of
    the effective resolvent} \\ \end{center}
\begin{algo}
\label{algo:diag_G1N}
Fix $\omega^2 \geq 0$ and $\eta > 0$, and denote
$B = \omega^2+\ri\eta-
A_\text{sys}-\Sigma^\eta_L(\omega)-\Sigma^\eta_R(\omega)$.
Set 
\[
F_1 = -B_{1,1}^{-1} B_{1,2}, \qquad g = F_1.
\]
Then,
\begin{enumerate}[\quad (1)] 
\item For $i=2,\dots,N-1$, compute
\[
F_{i} = -(B_{i,i}+B_{i,i-1}F_{i-1})^{-1} B_{i,i+1},
\]
and replace $g$ by $g F_i$;
\item Set $[G^\eta_\text{sys}(\omega)]_{1,N} = g \, (B_{N,N} + B_{N,N-1}F_{N-1})^{-1}$.
\end{enumerate}
\end{algo}
}
\end{figure}

An alternative linear scaling algorithm to compute the transmission is proposed 
in Ref.~\onlinecite{LKBJ97}. This algorithm also allows to compute
the density of states in $\text{O}(N_\text{layers} N_\text{at}^3)$ operations, 
but gives wrong transmissions for non small values of $\eta$
(in the results presented in this work, this occurs for even for
values $\eta/\omega_\text{max}^2 = 10^{-4}$ for one dimensional chains long
enough or when the variance of the mass disorder is large enough).


\section{Numerical implementation of the Boltzmann equation}
\label{sec:app_Boltz_num}

To simplify the notations, we drop the variable $\omega$ in the phonon
distributions (since it merely indexes independent Boltzmann equations).
To solve~(\ref{eq:Boltzmann_n}), it is convenient first to symmetrize
the problem by introducing $f_{j,\sigma} = v_j n_{j,\sigma}$,
so that (\ref{eq:Boltzmann_n}) can be recast as
\begin{equation}
  \label{eq:Boltzmann_f}
  \left ( \sigma \partial_t + \partial_x \right ) f_{j,\sigma}(x,t)
  = \sum_{(j',\sigma')} A_{(j,\sigma),(j',\sigma')} \, f_{j',\sigma'}(x,t),
\end{equation} 
where the matrix $A$ is symmetric. More precisely,
\[
A_{(j,\sigma),(j',\sigma')} = \sigma \frac{
  W_{(j,\sigma),(j',\sigma')} }{v_j v_{j'}}
\]
when $(j',\sigma') \not = (j,\sigma)$ and 
\[
A_{(j,\sigma),(j,\sigma)} = - \sigma \sum_{(j',\sigma') \not = (j,\sigma)} \frac{
  W_{(j,\sigma),(j',\sigma')} }{v_j \, v_{j'}}.
\]
The boundary conditions for (\ref{eq:Boltzmann_f}) are obtained from
the boundary conditions for (\ref{eq:Boltzmann_n}):
\[
f_{j,+}(0,t) = v_j, \qquad f_{j,-}(L,t) = 0.
\]
We denote $f=(f_{1,+},\dots,f_{N,+},f_{1,-},\dots,f_{N,-})$.
Recall that $2N$ is the number of conducting channels at this pulsation.
The stationary solution of~(\ref{eq:Boltzmann_f})
is such that $f(x) = \exp(xA) \, f(0)$, and thus $f(0) = \exp(-L A) \,
f(L)$, where we dropped the argument $t$. 
This expression has a meaning since $A$ is real symmetric hence
defines a self-adjoint operator.
Expressing $\exp(-L A)$ as a $2 \times 2$ matrix where
the entries are $N \times N$ matrices:
\[
\exp(-LA) = \left( \begin{array}{cc} 
B_{11}(L) & B_{12}(L) \\
B_{21}(L) & B_{22}(L) \\
\end{array} \right),
\]
it holds $B_{11}(L) \cdot (f_{1,+}(L),\dots,f_{N,+}(L))^t = (v_1,\dots,v_N)^t$.
Defining the $N \times N$ matrix ${\cal B}(L) = B_{11}^{-1}(L)$ when
this inverse exists, the transmission is finally
\[
T = N \frac{\dps \sum_{i=1}^N f_{i,+}(L)}{\dps \sum_{i=1}^N v_i} 
= N \frac{\dps \sum_{i=1}^N \sum_{j=1}^N {\cal B}_{ij}(L) v_j}{\dps
  \sum_{i=1}^N v_i}.
\]

In practice, it is not possible to use this formula, since the
matrix $B_{11}(L)$ becomes singular when $L$ is large. In those
cases, we compute a stationary solution of the Boltzmann 
equation~(\ref{eq:Boltzmann_f}) by integrating in time the
corresponding partial differential equation (PDE). The discretization
is done on a reference geometry (thanks to the hyperbolic scaling of
the transport part, it is possible to map a system of size $L$ to a
system of size unity, upon rescaling time and the scattering terms by a factor
$L$). The numerical scheme uses a
splitting of the dynamics into a transport part (which is implemented
with a simple upwind scheme\cite{GodlewskiRaviart} since the sign of the
velocities are known and are constant), and a collisional part (which
is integrated exactly using a matrix exponential). The stability
conditions on the time step (the so-called CFL conditions in the
literature on systems of conservation laws\cite{GodlewskiRaviart}) 
arise therefore only from
the transport part of the equation, and are easy to state since the
geometry is kept fixed.
We have checked that the algebraic method using the inversion of matrix
exponentials and the computation of the steady state through the time
integration of the PDE give the same results when the
condition number of the matrix $B_{11}$ is not too large. Another
consistency check is the fact that the computed transmission is almost
constant in space (up to the convergence error).

In the simulation results presented in this work, we have used $N_x = 200$
discretization points on the reference interval $[0,1]$, and a time step
$\Delta t=1.25 \times 10^{-3}$ for the time integration
of~(\ref{eq:Boltzmann_f}). 
We have checked the convergence of the results with respect to $N_x$
and $\Delta t$.

The scattering matrices~(\ref{eq:facteur_isotopic_disorder}) have been obtained by considering
a discrete phonon spectrum computed with $10^4$ $k$-points in the first
Brillouin zone, and by using windows of width $3.2$~cm$^{-1}$ centered around the pulsations 
$\omega$ of interest to list all the branches at this pulsation.
An approximation of the phonon velocity is obtained through finite
differences. Some scattering matrices are discarded in the end when
they contain too large values (which happens when the velocity of a
branch vanishes). This represents
less than 1\% of the cases for the profiles computed.
Such a procedure 
cannot however be used when the CNT index increases since more and more
branches start with a slope zero, and refined
strategies should then be employed for larger systems.


\section{Derivation of the Boltzmann formula~(\ref{eq:Boltz_1D})}
\label{sec:app_Boltz_1D}

For one dimensional chains, there is a single conducting branch
parametrized by $k$ in the first Brillouin zone:
\begin{equation}
  \label{eq:1D_boltz_steady_state}
  \omega(k)^2 = 2(1-\cos(ka)), \qquad -\frac{\pi}{a} \leq k <
  \frac{\pi}{a},
\end{equation}
where $a$ denotes the lattice spacing in this appendix (In
section~\ref{sec:one_dimensional_results}, $a=1$).
The steady-state Boltzmann equation at some frequency $\omega$ reduces to
\[
\partial_x n_+(x) = \frac{n_-(x)-n_+(x)}{l}, 
\]
\[
-\partial_x n_-(x) = \frac{n_+(x)-n_-(x)}{l},
\]
where the dependence on $\omega$ is not written out explicitly, and
\[
l^{-1} \equiv l(\omega)^{-1} = \frac{W_{+,-}(\omega)}{v^2(\omega)} = a 
\frac{\text{Var}(m)}{\langle m \rangle^2}
\frac{\omega^2(k)}{\omega'(k)^2}.
\]
From~(\ref{eq:1D_boltz_steady_state}), $\partial_x (n_+-n_-) =
0$. Using the boundary conditions $n_+(0) = 1$, $n_-(L) = 0$, and
denoting by $n_+(L) = T$ and $n_-(0) = R$ the proportions of
transmitted and reflected waves respectively, it follows $(n_+-n_-)(x) = (n_+-n_-)(0) =
(n_+-n_-)(L) = T$. Plugging this result
in~(\ref{eq:1D_boltz_steady_state}),
\[
\partial_x n_+(x) = -\frac{T}{l},
\]
so that 
\[
n_+(x) = 1 - \frac{x}{l} T.
\]
The transmission is then computed at $x=L$ using the above expression for $n_+$:
\[
T_L(\omega) = \frac{l(\omega)}{l(\omega) + L}.
\]
To obtain~(\ref{eq:Boltz_1D}), it remains to precise the value of $l^{-1}
\equiv l(\omega)^{-1}$.
From~(\ref{eq:1D_boltz_steady_state}), the phonon velocity is
\[
\omega'(k) = \frac{a \sin(ka)}{\omega},
\] 
so that the result follows using $\sin(ka)^2 = 1 - \cos(ka)^2 = 1 -
(1-\omega^2/2)^2 = \omega^2 (1-\omega^2/4)$.

%
%


\begin{thebibliography}{72}
\expandafter\ifx\csname natexlab\endcsname\relax\def\natexlab#1{#1}\fi
\expandafter\ifx\csname bibnamefont\endcsname\relax
  \def\bibnamefont#1{#1}\fi
\expandafter\ifx\csname bibfnamefont\endcsname\relax
  \def\bibfnamefont#1{#1}\fi
\expandafter\ifx\csname citenamefont\endcsname\relax
  \def\citenamefont#1{#1}\fi
\expandafter\ifx\csname url\endcsname\relax
  \def\url#1{\texttt{#1}}\fi
\expandafter\ifx\csname urlprefix\endcsname\relax\def\urlprefix{URL }\fi
\providecommand{\bibinfo}[2]{#2}
\providecommand{\eprint}[2][]{\url{#2}}

\bibitem[{\citenamefont{Yu et~al.}(2005)\citenamefont{Yu, Shi, Yao, Li, and
  Majumdar}}]{YSYLM05}
\bibinfo{author}{\bibfnamefont{C.}~\bibnamefont{Yu}},
  \bibinfo{author}{\bibfnamefont{L.}~\bibnamefont{Shi}},
  \bibinfo{author}{\bibfnamefont{Z.}~\bibnamefont{Yao}},
  \bibinfo{author}{\bibfnamefont{D.}~\bibnamefont{Li}}, \bibnamefont{and}
  \bibinfo{author}{\bibfnamefont{A.}~\bibnamefont{Majumdar}},
  \bibinfo{journal}{Nano Lett.} \textbf{\bibinfo{volume}{5}},
  \bibinfo{pages}{1842} (\bibinfo{year}{2005}).

\bibitem[{\citenamefont{Hone et~al.}(1999)\citenamefont{Hone, Whitney, Piskoti,
  and Zettl}}]{HWPZ99}
\bibinfo{author}{\bibfnamefont{J.}~\bibnamefont{Hone}},
  \bibinfo{author}{\bibfnamefont{M.}~\bibnamefont{Whitney}},
  \bibinfo{author}{\bibfnamefont{C.}~\bibnamefont{Piskoti}}, \bibnamefont{and}
  \bibinfo{author}{\bibfnamefont{A.}~\bibnamefont{Zettl}},
  \bibinfo{journal}{Phys. Rev. B} \textbf{\bibinfo{volume}{59}},
  \bibinfo{pages}{R2514} (\bibinfo{year}{1999}).

\bibitem[{\citenamefont{Yamamoto et~al.}(2007)\citenamefont{Yamamoto, Nakazawa,
  and Watanabe}}]{YNW07}
\bibinfo{author}{\bibfnamefont{T.}~\bibnamefont{Yamamoto}},
  \bibinfo{author}{\bibfnamefont{Y.}~\bibnamefont{Nakazawa}}, \bibnamefont{and}
  \bibinfo{author}{\bibfnamefont{K.}~\bibnamefont{Watanabe}},
  \bibinfo{journal}{New J. Phys.} \textbf{\bibinfo{volume}{9}},
  \bibinfo{pages}{245} (\bibinfo{year}{2007}).

\bibitem[{\citenamefont{Rieder et~al.}(1967)\citenamefont{Rieder, Lebowitz, and
  Lieb}}]{RLL67}
\bibinfo{author}{\bibfnamefont{Z.}~\bibnamefont{Rieder}},
  \bibinfo{author}{\bibfnamefont{J.}~\bibnamefont{Lebowitz}}, \bibnamefont{and}
  \bibinfo{author}{\bibfnamefont{E.}~\bibnamefont{Lieb}}, \bibinfo{journal}{J.
  Math. Phys.} \textbf{\bibinfo{volume}{8}}, \bibinfo{pages}{1073}
  (\bibinfo{year}{1967}).

\bibitem[{\citenamefont{Wang et~al.}(2007)\citenamefont{Wang, Tang, Zheng,
  Zhang, and Zhu}}]{WTZZZ07}
\bibinfo{author}{\bibfnamefont{Z.}~\bibnamefont{Wang}},
  \bibinfo{author}{\bibfnamefont{D.}~\bibnamefont{Tang}},
  \bibinfo{author}{\bibfnamefont{X.}~\bibnamefont{Zheng}},
  \bibinfo{author}{\bibfnamefont{W.}~\bibnamefont{Zhang}}, \bibnamefont{and}
  \bibinfo{author}{\bibfnamefont{Y.}~\bibnamefont{Zhu}},
  \bibinfo{journal}{Nanotechnology} \textbf{\bibinfo{volume}{18}},
  \bibinfo{pages}{475714} (\bibinfo{year}{2007}).

\bibitem[{\citenamefont{Chang et~al.}(2008)\citenamefont{Chang, Okawa, Garcia,
  Majumdar, and Zettl}}]{COGMZ08}
\bibinfo{author}{\bibfnamefont{C.~W.} \bibnamefont{Chang}},
  \bibinfo{author}{\bibfnamefont{D.}~\bibnamefont{Okawa}},
  \bibinfo{author}{\bibfnamefont{H.}~\bibnamefont{Garcia}},
  \bibinfo{author}{\bibfnamefont{A.}~\bibnamefont{Majumdar}}, \bibnamefont{and}
  \bibinfo{author}{\bibfnamefont{A.}~\bibnamefont{Zettl}},
  \bibinfo{journal}{Phys. Rev. Lett.} \textbf{\bibinfo{volume}{101}},
  \bibinfo{pages}{075903} (\bibinfo{year}{2008}).

\bibitem[{\citenamefont{Pop et~al.}(2006)\citenamefont{Pop, Mann, Wang,
  Goodson, and Dai}}]{PMWGD06}
\bibinfo{author}{\bibfnamefont{E.}~\bibnamefont{Pop}},
  \bibinfo{author}{\bibfnamefont{D.}~\bibnamefont{Mann}},
  \bibinfo{author}{\bibfnamefont{Q.}~\bibnamefont{Wang}},
  \bibinfo{author}{\bibfnamefont{K.}~\bibnamefont{Goodson}}, \bibnamefont{and}
  \bibinfo{author}{\bibfnamefont{H.}~\bibnamefont{Dai}}, \bibinfo{journal}{Nano
  Lett.} \textbf{\bibinfo{volume}{6}}, \bibinfo{pages}{96}
  (\bibinfo{year}{2006}).

\bibitem[{\citenamefont{Simon et~al.}(2005)\citenamefont{Simon, Kramberger,
  Pfeiffer, Kuzmany, Zolyomi, Kurti, Singer, and Alloul}}]{SKPKZKSA05}
\bibinfo{author}{\bibfnamefont{F.}~\bibnamefont{Simon}},
  \bibinfo{author}{\bibfnamefont{C.}~\bibnamefont{Kramberger}},
  \bibinfo{author}{\bibfnamefont{R.}~\bibnamefont{Pfeiffer}},
  \bibinfo{author}{\bibfnamefont{H.}~\bibnamefont{Kuzmany}},
  \bibinfo{author}{\bibfnamefont{V.}~\bibnamefont{Zolyomi}},
  \bibinfo{author}{\bibfnamefont{J.}~\bibnamefont{Kurti}},
  \bibinfo{author}{\bibfnamefont{P.}~\bibnamefont{Singer}}, \bibnamefont{and}
  \bibinfo{author}{\bibfnamefont{H.}~\bibnamefont{Alloul}},
  \bibinfo{journal}{Phys. Rev. Lett.} \textbf{\bibinfo{volume}{95}},
  \bibinfo{pages}{017401} (\bibinfo{year}{2005}).

\bibitem[{\citenamefont{Chang et~al.}(2006)\citenamefont{Chang, Fennimore,
  Afanasiev, Okawa, Ikuno, Garcia, Li, Majumdar, and Zettl}}]{CFAOIG06}
\bibinfo{author}{\bibfnamefont{C.}~\bibnamefont{Chang}},
  \bibinfo{author}{\bibfnamefont{A.}~\bibnamefont{Fennimore}},
  \bibinfo{author}{\bibfnamefont{A.}~\bibnamefont{Afanasiev}},
  \bibinfo{author}{\bibfnamefont{D.}~\bibnamefont{Okawa}},
  \bibinfo{author}{\bibfnamefont{I.}~\bibnamefont{Ikuno}},
  \bibinfo{author}{\bibfnamefont{H.}~\bibnamefont{Garcia}},
  \bibinfo{author}{\bibfnamefont{D.}~\bibnamefont{Li}},
  \bibinfo{author}{\bibfnamefont{A.}~\bibnamefont{Majumdar}}, \bibnamefont{and}
  \bibinfo{author}{\bibfnamefont{A.}~\bibnamefont{Zettl}},
  \bibinfo{journal}{Phys. Rev. Lett.} \textbf{\bibinfo{volume}{97}},
  \bibinfo{pages}{085901} (\bibinfo{year}{2006}).

\bibitem[{\citenamefont{Matsuda and Ishii}(1970)}]{MI70}
\bibinfo{author}{\bibfnamefont{H.}~\bibnamefont{Matsuda}} \bibnamefont{and}
  \bibinfo{author}{\bibfnamefont{K.}~\bibnamefont{Ishii}},
  \bibinfo{journal}{Suppl. Prog. Theor. Phys.} \textbf{\bibinfo{volume}{45}},
  \bibinfo{pages}{56} (\bibinfo{year}{1970}).

\bibitem[{\citenamefont{Bonetto et~al.}(2000)\citenamefont{Bonetto, Lebowitz,
  and Rey-Bellet}}]{BLRB00}
\bibinfo{author}{\bibfnamefont{F.}~\bibnamefont{Bonetto}},
  \bibinfo{author}{\bibfnamefont{J.}~\bibnamefont{Lebowitz}}, \bibnamefont{and}
  \bibinfo{author}{\bibfnamefont{L.}~\bibnamefont{Rey-Bellet}}, in
  \emph{\bibinfo{booktitle}{Mathematical Physics 2000}}, edited by
  \bibinfo{editor}{\bibfnamefont{A.}~\bibnamefont{Fokas}},
  \bibinfo{editor}{\bibfnamefont{A.}~\bibnamefont{Grigoryan}},
  \bibinfo{editor}{\bibfnamefont{T.}~\bibnamefont{Kibble}}, \bibnamefont{and}
  \bibinfo{editor}{\bibfnamefont{B.}~\bibnamefont{Zegarlinsky}}
  (\bibinfo{publisher}{Imperial College Press}, \bibinfo{year}{2000}), pp.
  \bibinfo{pages}{128--151}.

\bibitem[{\citenamefont{Lepri et~al.}(2003{\natexlab{a}})\citenamefont{Lepri,
  Livi, and Politi}}]{LLP03}
\bibinfo{author}{\bibfnamefont{S.}~\bibnamefont{Lepri}},
  \bibinfo{author}{\bibfnamefont{R.}~\bibnamefont{Livi}}, \bibnamefont{and}
  \bibinfo{author}{\bibfnamefont{A.}~\bibnamefont{Politi}},
  \bibinfo{journal}{Phys. Rep.} \textbf{\bibinfo{volume}{377}},
  \bibinfo{pages}{1} (\bibinfo{year}{2003}{\natexlab{a}}).

\bibitem[{\citenamefont{Basile et~al.}(2006)\citenamefont{Basile, Bernardin,
  and Olla}}]{BBO06}
\bibinfo{author}{\bibfnamefont{G.}~\bibnamefont{Basile}},
  \bibinfo{author}{\bibfnamefont{C.}~\bibnamefont{Bernardin}},
  \bibnamefont{and} \bibinfo{author}{\bibfnamefont{S.}~\bibnamefont{Olla}},
  \bibinfo{journal}{Phys. Rev. Lett.} \textbf{\bibinfo{volume}{96}},
  \bibinfo{pages}{204303} (\bibinfo{year}{2006}).

\bibitem[{\citenamefont{Bolsterli et~al.}(1970)\citenamefont{Bolsterli, Rich,
  and Visscher}}]{BRV70}
\bibinfo{author}{\bibfnamefont{M.}~\bibnamefont{Bolsterli}},
  \bibinfo{author}{\bibfnamefont{M.}~\bibnamefont{Rich}}, \bibnamefont{and}
  \bibinfo{author}{\bibfnamefont{W.}~\bibnamefont{Visscher}},
  \bibinfo{journal}{Phys. Rev. A} \textbf{\bibinfo{volume}{1}},
  \bibinfo{pages}{1086} (\bibinfo{year}{1970}).

\bibitem[{\citenamefont{Bonetto et~al.}(2004)\citenamefont{Bonetto, Lebowitz,
  and Lukkarinen}}]{BLL04}
\bibinfo{author}{\bibfnamefont{F.}~\bibnamefont{Bonetto}},
  \bibinfo{author}{\bibfnamefont{J.}~\bibnamefont{Lebowitz}}, \bibnamefont{and}
  \bibinfo{author}{\bibfnamefont{J.}~\bibnamefont{Lukkarinen}},
  \bibinfo{journal}{J. Stat. Phys.} \textbf{\bibinfo{volume}{116}},
  \bibinfo{pages}{783} (\bibinfo{year}{2004}).

\bibitem[{\citenamefont{Narayan and Ramaswamy}(2002)}]{NS02}
\bibinfo{author}{\bibfnamefont{O.}~\bibnamefont{Narayan}} \bibnamefont{and}
  \bibinfo{author}{\bibfnamefont{S.}~\bibnamefont{Ramaswamy}},
  \bibinfo{journal}{Phys. Rev. Lett.} \textbf{\bibinfo{volume}{89}},
  \bibinfo{pages}{200601} (\bibinfo{year}{2002}).

\bibitem[{\citenamefont{Lepri et~al.}(2003{\natexlab{b}})\citenamefont{Lepri,
  Livi, and Politi}}]{LLP03b}
\bibinfo{author}{\bibfnamefont{S.}~\bibnamefont{Lepri}},
  \bibinfo{author}{\bibfnamefont{R.}~\bibnamefont{Livi}}, \bibnamefont{and}
  \bibinfo{author}{\bibfnamefont{A.}~\bibnamefont{Politi}},
  \bibinfo{journal}{Phys. Rev. E} \textbf{\bibinfo{volume}{68}},
  \bibinfo{pages}{067102} (\bibinfo{year}{2003}{\natexlab{b}}).

\bibitem[{\citenamefont{Mai et~al.}(2007)\citenamefont{Mai, Dhar, and
  Narayan}}]{MDN07}
\bibinfo{author}{\bibfnamefont{T.}~\bibnamefont{Mai}},
  \bibinfo{author}{\bibfnamefont{A.}~\bibnamefont{Dhar}}, \bibnamefont{and}
  \bibinfo{author}{\bibfnamefont{O.}~\bibnamefont{Narayan}},
  \bibinfo{journal}{Phys. Rev. Lett.} \textbf{\bibinfo{volume}{98}},
  \bibinfo{pages}{184301} (\bibinfo{year}{2007}).

\bibitem[{\citenamefont{Dhar}(2001)}]{Dhar01}
\bibinfo{author}{\bibfnamefont{A.}~\bibnamefont{Dhar}}, \bibinfo{journal}{Phys.
  Rev. Lett.} \textbf{\bibinfo{volume}{86}}, \bibinfo{pages}{5882}
  (\bibinfo{year}{2001}).

\bibitem[{\citenamefont{Yamamoto and Watanabe}(2006)}]{YW06}
\bibinfo{author}{\bibfnamefont{T.}~\bibnamefont{Yamamoto}} \bibnamefont{and}
  \bibinfo{author}{\bibfnamefont{K.}~\bibnamefont{Watanabe}},
  \bibinfo{journal}{Phys. Rev. Lett.} \textbf{\bibinfo{volume}{96}},
  \bibinfo{pages}{255503} (\bibinfo{year}{2006}).

\bibitem[{\citenamefont{Evans}(1982)}]{Evans82}
\bibinfo{author}{\bibfnamefont{D.}~\bibnamefont{Evans}},
  \bibinfo{journal}{Phys. Lett. A} \textbf{\bibinfo{volume}{91}},
  \bibinfo{pages}{457} (\bibinfo{year}{1982}).

\bibitem[{\citenamefont{Yao et~al.}(2005)\citenamefont{Yao, Wang, Li, and
  Liu}}]{YWLL05}
\bibinfo{author}{\bibfnamefont{Z.}~\bibnamefont{Yao}},
  \bibinfo{author}{\bibfnamefont{J.}~\bibnamefont{Wang}},
  \bibinfo{author}{\bibfnamefont{B.}~\bibnamefont{Li}}, \bibnamefont{and}
  \bibinfo{author}{\bibfnamefont{G.}~\bibnamefont{Liu}},
  \bibinfo{journal}{Phys. Rev. B} \textbf{\bibinfo{volume}{71}},
  \bibinfo{pages}{085417} (\bibinfo{year}{2005}).

\bibitem[{\citenamefont{Mingo and Broido}(2005{\natexlab{a}})}]{MB05}
\bibinfo{author}{\bibfnamefont{N.}~\bibnamefont{Mingo}} \bibnamefont{and}
  \bibinfo{author}{\bibfnamefont{D.}~\bibnamefont{Broido}},
  \bibinfo{journal}{Phys. Rev. Lett.} \textbf{\bibinfo{volume}{95}},
  \bibinfo{pages}{096105} (\bibinfo{year}{2005}{\natexlab{a}}).

\bibitem[{\citenamefont{Lukes and Zhong}(2007)}]{LZ07}
\bibinfo{author}{\bibfnamefont{J.}~\bibnamefont{Lukes}} \bibnamefont{and}
  \bibinfo{author}{\bibfnamefont{H.}~\bibnamefont{Zhong}}, \bibinfo{journal}{J.
  Heat Transfer} \textbf{\bibinfo{volume}{129}}, \bibinfo{pages}{705}
  (\bibinfo{year}{2007}).

\bibitem[{\citenamefont{Maruyama et~al.}(2006)\citenamefont{Maruyama, Igarashi,
  Taniguchi, and Shiomi}}]{MITS06}
\bibinfo{author}{\bibfnamefont{S.}~\bibnamefont{Maruyama}},
  \bibinfo{author}{\bibfnamefont{Y.}~\bibnamefont{Igarashi}},
  \bibinfo{author}{\bibfnamefont{Y.}~\bibnamefont{Taniguchi}},
  \bibnamefont{and} \bibinfo{author}{\bibfnamefont{J.}~\bibnamefont{Shiomi}},
  \bibinfo{journal}{Journal of thermal science and technology}
  \textbf{\bibinfo{volume}{1}}, \bibinfo{pages}{138} (\bibinfo{year}{2006}).

\bibitem[{\citenamefont{Maruyama}(2002)}]{Maruyama02}
\bibinfo{author}{\bibfnamefont{S.}~\bibnamefont{Maruyama}},
  \bibinfo{journal}{Physica B} \textbf{\bibinfo{volume}{323}},
  \bibinfo{pages}{193} (\bibinfo{year}{2002}).

\bibitem[{\citenamefont{Zhang and Li}(2005)}]{ZL05}
\bibinfo{author}{\bibfnamefont{G.}~\bibnamefont{Zhang}} \bibnamefont{and}
  \bibinfo{author}{\bibfnamefont{B.~W.} \bibnamefont{Li}}, \bibinfo{journal}{J.
  Chem. Phys.} \textbf{\bibinfo{volume}{123}}, \bibinfo{pages}{114714}
  (\bibinfo{year}{2005}).

\bibitem[{\citenamefont{Pan et~al.}(2007)\citenamefont{Pan, Xu, and
  Zhu}}]{PXZ07}
\bibinfo{author}{\bibfnamefont{R.}~\bibnamefont{Pan}},
  \bibinfo{author}{\bibfnamefont{Z.}~\bibnamefont{Xu}}, \bibnamefont{and}
  \bibinfo{author}{\bibfnamefont{Z.}~\bibnamefont{Zhu}},
  \bibinfo{journal}{Chinese Phys. Lett.} \textbf{\bibinfo{volume}{24}},
  \bibinfo{pages}{1321} (\bibinfo{year}{2007}).

\bibitem[{\citenamefont{Shiomi and Maruyama}(2006)}]{SM06}
\bibinfo{author}{\bibfnamefont{J.}~\bibnamefont{Shiomi}} \bibnamefont{and}
  \bibinfo{author}{\bibfnamefont{S.}~\bibnamefont{Maruyama}},
  \bibinfo{journal}{Phys. Rev. B} \textbf{\bibinfo{volume}{74}},
  \bibinfo{pages}{155401} (\bibinfo{year}{2006}).

\bibitem[{\citenamefont{Bi et~al.}(2006)\citenamefont{Bi, Chen, Yang, Wang, and
  Chen}}]{BCYWC06}
\bibinfo{author}{\bibfnamefont{K.}~\bibnamefont{Bi}},
  \bibinfo{author}{\bibfnamefont{Y.}~\bibnamefont{Chen}},
  \bibinfo{author}{\bibfnamefont{J.}~\bibnamefont{Yang}},
  \bibinfo{author}{\bibfnamefont{Y.}~\bibnamefont{Wang}}, \bibnamefont{and}
  \bibinfo{author}{\bibfnamefont{M.}~\bibnamefont{Chen}},
  \bibinfo{journal}{Phys. Lett. A} \textbf{\bibinfo{volume}{350}},
  \bibinfo{pages}{150} (\bibinfo{year}{2006}).

\bibitem[{\citenamefont{Shenogin et~al.}(2004)\citenamefont{Shenogin, Bodapati,
  Xue, Ozisik, and Keblinski}}]{SBXOK04}
\bibinfo{author}{\bibfnamefont{S.}~\bibnamefont{Shenogin}},
  \bibinfo{author}{\bibfnamefont{A.}~\bibnamefont{Bodapati}},
  \bibinfo{author}{\bibfnamefont{L.}~\bibnamefont{Xue}},
  \bibinfo{author}{\bibfnamefont{R.}~\bibnamefont{Ozisik}}, \bibnamefont{and}
  \bibinfo{author}{\bibfnamefont{P.}~\bibnamefont{Keblinski}},
  \bibinfo{journal}{Appl. Phys. Lett.} \textbf{\bibinfo{volume}{85}},
  \bibinfo{pages}{2229} (\bibinfo{year}{2004}).

\bibitem[{\citenamefont{Zhang et~al.}(2007)\citenamefont{Zhang, Fisher, and
  Mingo}}]{ZFM07}
\bibinfo{author}{\bibfnamefont{W.}~\bibnamefont{Zhang}},
  \bibinfo{author}{\bibfnamefont{T.}~\bibnamefont{Fisher}}, \bibnamefont{and}
  \bibinfo{author}{\bibfnamefont{N.}~\bibnamefont{Mingo}},
  \bibinfo{journal}{Numer. Heat Tr. B - Fund.} \textbf{\bibinfo{volume}{51}},
  \bibinfo{pages}{333} (\bibinfo{year}{2007}).

\bibitem[{\citenamefont{Datta}(2005)}]{Datta2006}
\bibinfo{author}{\bibfnamefont{S.}~\bibnamefont{Datta}},
  \emph{\bibinfo{title}{Quantum {Transport: From Atom to Transistor}}}
  (\bibinfo{publisher}{Cambridge University Press}, \bibinfo{year}{2005}).

\bibitem[{\citenamefont{Datta}(2000)}]{Datta2000}
\bibinfo{author}{\bibfnamefont{S.}~\bibnamefont{Datta}},
  \bibinfo{journal}{Superlattices and microsctructures}
  \textbf{\bibinfo{volume}{28}}, \bibinfo{pages}{253} (\bibinfo{year}{2000}).

\bibitem[{\citenamefont{Ford et~al.}(1965)\citenamefont{Ford, Kac, and
  Mazur}}]{FKM65}
\bibinfo{author}{\bibfnamefont{G.}~\bibnamefont{Ford}},
  \bibinfo{author}{\bibfnamefont{M.}~\bibnamefont{Kac}}, \bibnamefont{and}
  \bibinfo{author}{\bibfnamefont{P.}~\bibnamefont{Mazur}}, \bibinfo{journal}{J.
  Math. Phys.} \textbf{\bibinfo{volume}{6}}, \bibinfo{pages}{504}
  (\bibinfo{year}{1965}).

\bibitem[{\citenamefont{Dhar and Roy}(2006)}]{DR06}
\bibinfo{author}{\bibfnamefont{A.}~\bibnamefont{Dhar}} \bibnamefont{and}
  \bibinfo{author}{\bibfnamefont{D.}~\bibnamefont{Roy}}, \bibinfo{journal}{J.
  Stat. Phys.} \textbf{\bibinfo{volume}{125}}, \bibinfo{pages}{805}
  (\bibinfo{year}{2006}).

\bibitem[{\citenamefont{Aschbacher et~al.}(2007)\citenamefont{Aschbacher,
  Jaksic, Pautrat, and Pillet}}]{AJPP07}
\bibinfo{author}{\bibfnamefont{W.}~\bibnamefont{Aschbacher}},
  \bibinfo{author}{\bibfnamefont{V.}~\bibnamefont{Jaksic}},
  \bibinfo{author}{\bibfnamefont{Y.}~\bibnamefont{Pautrat}}, \bibnamefont{and}
  \bibinfo{author}{\bibfnamefont{C.-A.} \bibnamefont{Pillet}},
  \bibinfo{journal}{J. Math. Phys.} \textbf{\bibinfo{volume}{48}},
  \bibinfo{pages}{032101} (\bibinfo{year}{2007}).

\bibitem[{\citenamefont{Ashcroft and Mermin}(1976)}]{AshcroftMermin}
\bibinfo{author}{\bibfnamefont{N.}~\bibnamefont{Ashcroft}} \bibnamefont{and}
  \bibinfo{author}{\bibfnamefont{N.}~\bibnamefont{Mermin}},
  \emph{\bibinfo{title}{Solid State Physics}} (\bibinfo{publisher}{Saunders
  College Publishing}, \bibinfo{year}{1976}).

\bibitem[{\citenamefont{Baroni et~al.}(2001)\citenamefont{Baroni, de~Gironcoli,
  and Corso}}]{BdGdC01}
\bibinfo{author}{\bibfnamefont{S.}~\bibnamefont{Baroni}},
  \bibinfo{author}{\bibfnamefont{S.}~\bibnamefont{de~Gironcoli}},
  \bibnamefont{and} \bibinfo{author}{\bibfnamefont{A.~D.} \bibnamefont{Corso}},
  \bibinfo{journal}{Rev. Mod. Phys.} \textbf{\bibinfo{volume}{73}},
  \bibinfo{pages}{515} (\bibinfo{year}{2001}).

\bibitem[{\citenamefont{Jaksic}(2006)}]{Jaksic}
\bibinfo{author}{\bibfnamefont{V.}~\bibnamefont{Jaksic}},
  \bibinfo{journal}{Lecture Notes in Mathematics}
  \textbf{\bibinfo{volume}{1880}}, \bibinfo{pages}{235} (\bibinfo{year}{2006}).

\bibitem[{\citenamefont{Rego and Kirczenow}(1998)}]{RK98}
\bibinfo{author}{\bibfnamefont{L.}~\bibnamefont{Rego}} \bibnamefont{and}
  \bibinfo{author}{\bibfnamefont{G.}~\bibnamefont{Kirczenow}},
  \bibinfo{journal}{Phys. Rev. Lett.} \textbf{\bibinfo{volume}{81}},
  \bibinfo{pages}{232} (\bibinfo{year}{1998}).

\bibitem[{\citenamefont{Schwab et~al.}(2000)\citenamefont{Schwab, Henriksen,
  Worlock, and Roukes}}]{SHWR00}
\bibinfo{author}{\bibfnamefont{K.}~\bibnamefont{Schwab}},
  \bibinfo{author}{\bibfnamefont{E.}~\bibnamefont{Henriksen}},
  \bibinfo{author}{\bibfnamefont{J.}~\bibnamefont{Worlock}}, \bibnamefont{and}
  \bibinfo{author}{\bibfnamefont{M.}~\bibnamefont{Roukes}},
  \bibinfo{journal}{Nature} \textbf{\bibinfo{volume}{404}},
  \bibinfo{pages}{974} (\bibinfo{year}{2000}).

\bibitem[{\citenamefont{Mingo and Broido}(2005{\natexlab{b}})}]{MB05_bis}
\bibinfo{author}{\bibfnamefont{N.}~\bibnamefont{Mingo}} \bibnamefont{and}
  \bibinfo{author}{\bibfnamefont{D.}~\bibnamefont{Broido}},
  \bibinfo{journal}{Nano Lett.} \textbf{\bibinfo{volume}{5}},
  \bibinfo{pages}{1221} (\bibinfo{year}{2005}{\natexlab{b}}).

\bibitem[{\citenamefont{Spohn}(2006)}]{Spohn06}
\bibinfo{author}{\bibfnamefont{H.}~\bibnamefont{Spohn}}, \bibinfo{journal}{J.
  Stat. Phys.} \textbf{\bibinfo{volume}{124}}, \bibinfo{pages}{1041}
  (\bibinfo{year}{2006}).

\bibitem[{\citenamefont{Lukkarinen and Spohn}(2007)}]{SL07}
\bibinfo{author}{\bibfnamefont{J.}~\bibnamefont{Lukkarinen}} \bibnamefont{and}
  \bibinfo{author}{\bibfnamefont{H.}~\bibnamefont{Spohn}},
  \bibinfo{journal}{Arch. Rational Mech. Anal.} \textbf{\bibinfo{volume}{183}},
  \bibinfo{pages}{93} (\bibinfo{year}{2007}).

\bibitem[{\citenamefont{Tamura}(1983)}]{Tamura83}
\bibinfo{author}{\bibfnamefont{S.}~\bibnamefont{Tamura}},
  \bibinfo{journal}{Phys. Rev. B} \textbf{\bibinfo{volume}{27}},
  \bibinfo{pages}{858} (\bibinfo{year}{1983}).

\bibitem[{\citenamefont{Widulle et~al.}(2002)\citenamefont{Widulle, Serrano,
  and Cardona}}]{WSC02}
\bibinfo{author}{\bibfnamefont{F.}~\bibnamefont{Widulle}},
  \bibinfo{author}{\bibfnamefont{J.}~\bibnamefont{Serrano}}, \bibnamefont{and}
  \bibinfo{author}{\bibfnamefont{M.}~\bibnamefont{Cardona}},
  \bibinfo{journal}{Phys. Rev. B} \textbf{\bibinfo{volume}{65}},
  \bibinfo{pages}{075206} (\bibinfo{year}{2002}).

\bibitem[{\citenamefont{Vandecasteel et~al.}(2008)\citenamefont{Vandecasteel,
  Lazzeri, and Mauri}}]{Niels}
\bibinfo{author}{\bibfnamefont{N.}~\bibnamefont{Vandecasteel}},
  \bibinfo{author}{\bibfnamefont{M.}~\bibnamefont{Lazzeri}}, \bibnamefont{and}
  \bibinfo{author}{\bibfnamefont{F.}~\bibnamefont{Mauri}},
  \bibinfo{journal}{unpublished}  (\bibinfo{year}{2008}).

\bibitem[{\citenamefont{Rubin and Greer}(1971)}]{RG71}
\bibinfo{author}{\bibfnamefont{R.}~\bibnamefont{Rubin}} \bibnamefont{and}
  \bibinfo{author}{\bibfnamefont{W.}~\bibnamefont{Greer}}, \bibinfo{journal}{J.
  Math. Phys.} \textbf{\bibinfo{volume}{12}}, \bibinfo{pages}{1686}
  (\bibinfo{year}{1971}).

\bibitem[{\citenamefont{Casher and Lebowitz}(1971)}]{CL71}
\bibinfo{author}{\bibfnamefont{A.}~\bibnamefont{Casher}} \bibnamefont{and}
  \bibinfo{author}{\bibfnamefont{J.}~\bibnamefont{Lebowitz}},
  \bibinfo{journal}{J. Math. Phys.} \textbf{\bibinfo{volume}{12}},
  \bibinfo{pages}{1701} (\bibinfo{year}{1971}).

\bibitem[{\citenamefont{Ishii}(1973)}]{Ishii73}
\bibinfo{author}{\bibfnamefont{K.}~\bibnamefont{Ishii}},
  \bibinfo{journal}{Suppl. Prog. Theor. Phys.} \textbf{\bibinfo{volume}{45}},
  \bibinfo{pages}{77} (\bibinfo{year}{1973}).

\bibitem[{\citenamefont{O'Connor and Lebowitz}(1974)}]{OcL74}
\bibinfo{author}{\bibfnamefont{A.}~\bibnamefont{O'Connor}} \bibnamefont{and}
  \bibinfo{author}{\bibfnamefont{J.}~\bibnamefont{Lebowitz}},
  \bibinfo{journal}{J. Math. Phys.} \textbf{\bibinfo{volume}{15}},
  \bibinfo{pages}{692} (\bibinfo{year}{1974}).

\bibitem[{\citenamefont{Keller et~al.}(1978)\citenamefont{Keller, Papanicolaou,
  and Weilenmann}}]{KPW78}
\bibinfo{author}{\bibfnamefont{J.}~\bibnamefont{Keller}},
  \bibinfo{author}{\bibfnamefont{G.}~\bibnamefont{Papanicolaou}},
  \bibnamefont{and}
  \bibinfo{author}{\bibfnamefont{J.}~\bibnamefont{Weilenmann}},
  \bibinfo{journal}{Commmun. Pure Appl. Math.} \textbf{\bibinfo{volume}{32}},
  \bibinfo{pages}{583} (\bibinfo{year}{1978}).

\bibitem[{\citenamefont{Rubin}(1968)}]{Rubin68}
\bibinfo{author}{\bibfnamefont{R.}~\bibnamefont{Rubin}}, \bibinfo{journal}{J.
  Math. Phys.} \textbf{\bibinfo{volume}{9}}, \bibinfo{pages}{2252}
  (\bibinfo{year}{1968}).

\bibitem[{\citenamefont{Furstenberg}(1963)}]{Furstenberg63}
\bibinfo{author}{\bibfnamefont{H.}~\bibnamefont{Furstenberg}},
  \bibinfo{journal}{Trans. Am. Math. Soc.} \textbf{\bibinfo{volume}{108}},
  \bibinfo{pages}{377} (\bibinfo{year}{1963}).

\bibitem[{\citenamefont{Economou}(2006)}]{Economou}
\bibinfo{author}{\bibfnamefont{E.}~\bibnamefont{Economou}},
  \emph{\bibinfo{title}{{Green's Function Methods in Quantum Physics (3rd
  Ed.)}}} (\bibinfo{publisher}{Springer}, \bibinfo{year}{2006}).

\bibitem[{\citenamefont{Avriller et~al.}(2007)\citenamefont{Avriller, Roche,
  Triozon, Blase, and Latil}}]{ARTBL07}
\bibinfo{author}{\bibfnamefont{R.}~\bibnamefont{Avriller}},
  \bibinfo{author}{\bibfnamefont{S.}~\bibnamefont{Roche}},
  \bibinfo{author}{\bibfnamefont{F.}~\bibnamefont{Triozon}},
  \bibinfo{author}{\bibfnamefont{X.}~\bibnamefont{Blase}}, \bibnamefont{and}
  \bibinfo{author}{\bibfnamefont{S.}~\bibnamefont{Latil}},
  \bibinfo{journal}{Modern Physics Letters B} \textbf{\bibinfo{volume}{21}},
  \bibinfo{pages}{1955} (\bibinfo{year}{2007}).

\bibitem[{\citenamefont{Likhachev et~al.}(2006)\citenamefont{Likhachev,
  Vinogradov, Astakhova, and Yakovenko}}]{LVAY06}
\bibinfo{author}{\bibfnamefont{V.}~\bibnamefont{Likhachev}},
  \bibinfo{author}{\bibfnamefont{G.}~\bibnamefont{Vinogradov}},
  \bibinfo{author}{\bibfnamefont{T.}~\bibnamefont{Astakhova}},
  \bibnamefont{and}
  \bibinfo{author}{\bibfnamefont{A.}~\bibnamefont{Yakovenko}},
  \bibinfo{journal}{Phys. Rev. B} \textbf{\bibinfo{volume}{73}},
  \bibinfo{pages}{016701} (\bibinfo{year}{2006}).

\bibitem[{\citenamefont{et~al.}()}]{QuantumEspresso}
\bibinfo{author}{\bibfnamefont{S.~B.} \bibnamefont{et~al.}},
  \bibinfo{journal}{http://www.quantum-espresso.org}  (????).

\bibitem[{\citenamefont{Saito et~al.}(1998)\citenamefont{Saito, Dresselhaus,
  and Dresselhaus}}]{SDD98}
\bibinfo{author}{\bibfnamefont{R.}~\bibnamefont{Saito}},
  \bibinfo{author}{\bibfnamefont{M.}~\bibnamefont{Dresselhaus}},
  \bibnamefont{and}
  \bibinfo{author}{\bibfnamefont{G.}~\bibnamefont{Dresselhaus}},
  \emph{\bibinfo{title}{Physical properties of carbon nanotubes}}
  (\bibinfo{publisher}{Imperial College Press}, \bibinfo{year}{1998}).

\bibitem[{\citenamefont{Mounet}(2005)}]{Mounet05}
\bibinfo{author}{\bibfnamefont{N.}~\bibnamefont{Mounet}},
  \emph{\bibinfo{title}{Structural, vibrational and thermodynamic properties of
  carbon allotropes from first-principles: diamond, graphite, and nanotubes}}
  (\bibinfo{publisher}{Master Thesis}, \bibinfo{address}{MIT},
  \bibinfo{year}{2005}).

\bibitem[{\citenamefont{Anderson}(1958)}]{Anderson58}
\bibinfo{author}{\bibfnamefont{P.}~\bibnamefont{Anderson}},
  \bibinfo{journal}{Phys. Rev.} \textbf{\bibinfo{volume}{109}},
  \bibinfo{pages}{1492} (\bibinfo{year}{1958}).

\bibitem[{\citenamefont{Mingo et~al.}(2008)\citenamefont{Mingo, Stewart,
  Broido, and Srivastava}}]{MSBS08}
\bibinfo{author}{\bibfnamefont{N.}~\bibnamefont{Mingo}},
  \bibinfo{author}{\bibfnamefont{D.~A.} \bibnamefont{Stewart}},
  \bibinfo{author}{\bibfnamefont{D.~A.} \bibnamefont{Broido}},
  \bibnamefont{and}
  \bibinfo{author}{\bibfnamefont{D.}~\bibnamefont{Srivastava}},
  \bibinfo{journal}{Phys. Rev. B} \textbf{\bibinfo{volume}{77}},
  \bibinfo{pages}{033418} (\bibinfo{year}{2008}).

\bibitem[{\citenamefont{da~Silva and Koiller}(1981)}]{GdSK81}
\bibinfo{author}{\bibfnamefont{C.~G.} \bibnamefont{da~Silva}} \bibnamefont{and}
  \bibinfo{author}{\bibfnamefont{B.}~\bibnamefont{Koiller}},
  \bibinfo{journal}{Solid State Commun.} \textbf{\bibinfo{volume}{40}},
  \bibinfo{pages}{215} (\bibinfo{year}{1981}).

\bibitem[{\citenamefont{Guinea et~al.}(1983)\citenamefont{Guinea, Tejedor,
  Flores, and Louis}}]{GTFL83}
\bibinfo{author}{\bibfnamefont{F.}~\bibnamefont{Guinea}},
  \bibinfo{author}{\bibfnamefont{C.}~\bibnamefont{Tejedor}},
  \bibinfo{author}{\bibfnamefont{F.}~\bibnamefont{Flores}}, \bibnamefont{and}
  \bibinfo{author}{\bibfnamefont{E.}~\bibnamefont{Louis}},
  \bibinfo{journal}{Phys. Rev. B} \textbf{\bibinfo{volume}{28}},
  \bibinfo{pages}{4397} (\bibinfo{year}{1983}).

\bibitem[{\citenamefont{Lopez-Sancho et~al.}(1984)\citenamefont{Lopez-Sancho,
  Lopez-Sancho, and Rubio}}]{LLR84}
\bibinfo{author}{\bibfnamefont{M.}~\bibnamefont{Lopez-Sancho}},
  \bibinfo{author}{\bibfnamefont{J.}~\bibnamefont{Lopez-Sancho}},
  \bibnamefont{and} \bibinfo{author}{\bibfnamefont{J.}~\bibnamefont{Rubio}},
  \bibinfo{journal}{J. Phys. F: Met. Phys.} \textbf{\bibinfo{volume}{14}},
  \bibinfo{pages}{1205} (\bibinfo{year}{1984}).

\bibitem[{\citenamefont{Lopez-Sancho et~al.}(1985)\citenamefont{Lopez-Sancho,
  Lopez-Sancho, and Rubio}}]{LLR85}
\bibinfo{author}{\bibfnamefont{M.}~\bibnamefont{Lopez-Sancho}},
  \bibinfo{author}{\bibfnamefont{J.}~\bibnamefont{Lopez-Sancho}},
  \bibnamefont{and} \bibinfo{author}{\bibfnamefont{J.}~\bibnamefont{Rubio}},
  \bibinfo{journal}{J. Phys. F: Met. Phys.} \textbf{\bibinfo{volume}{15}},
  \bibinfo{pages}{851} (\bibinfo{year}{1985}).

\bibitem[{\citenamefont{Lake et~al.}(1997)\citenamefont{Lake, Klimeck, Bowen,
  and Jovanovic}}]{LKBJ97}
\bibinfo{author}{\bibfnamefont{R.}~\bibnamefont{Lake}},
  \bibinfo{author}{\bibfnamefont{G.}~\bibnamefont{Klimeck}},
  \bibinfo{author}{\bibfnamefont{R.}~\bibnamefont{Bowen}}, \bibnamefont{and}
  \bibinfo{author}{\bibfnamefont{D.}~\bibnamefont{Jovanovic}},
  \bibinfo{journal}{J. Appl. Phys.} \textbf{\bibinfo{volume}{81}},
  \bibinfo{pages}{7845} (\bibinfo{year}{1997}).

\bibitem[{\citenamefont{Godlewski and Raviart}(1996)}]{GodlewskiRaviart}
\bibinfo{author}{\bibfnamefont{E.}~\bibnamefont{Godlewski}} \bibnamefont{and}
  \bibinfo{author}{\bibfnamefont{P.-A.} \bibnamefont{Raviart}},
  \emph{\bibinfo{title}{Numerical approximation of hyperbolic systems of
  conservation laws}}, vol. \bibinfo{volume}{118} of
  \emph{\bibinfo{series}{Applied mathematical sciences}}
  (\bibinfo{publisher}{Springer-Verlag}, \bibinfo{year}{1996}).

\bibitem[{\citenamefont{Ceperley and Alder}(1980)}]{LDA}
\bibinfo{author}{\bibfnamefont{D.~M.} \bibnamefont{Ceperley}} \bibnamefont{and}
  \bibinfo{author}{\bibfnamefont{B.~J.} \bibnamefont{Alder}},
  \bibinfo{journal}{Phys. Rev. Lett.} \textbf{\bibinfo{volume}{45}},
  \bibinfo{pages}{566} (\bibinfo{year}{1980}).

\bibitem[{\citenamefont{Troullier and Martins}(1991)}]{pseudo}
\bibinfo{author}{\bibfnamefont{N.}~\bibnamefont{Troullier}} \bibnamefont{and}
  \bibinfo{author}{\bibfnamefont{J.~L.} \bibnamefont{Martins}},
  \bibinfo{journal}{Phys. Rev. B} \textbf{\bibinfo{volume}{43}},
  \bibinfo{pages}{1993} (\bibinfo{year}{1991}).

\bibitem[{\citenamefont{Fuchs and Scheffler}(1999)}]{pseudo2}
\bibinfo{author}{\bibfnamefont{M.}~\bibnamefont{Fuchs}} \bibnamefont{and}
  \bibinfo{author}{\bibfnamefont{M.}~\bibnamefont{Scheffler}},
  \bibinfo{journal}{Comput. Phys. Commun.} \textbf{\bibinfo{volume}{119}},
  \bibinfo{pages}{67} (\bibinfo{year}{1999}).

\end{thebibliography}

\end{document}